\documentstyle[11pt,epsfig,graphicx]{article}
 
\newcommand \beq{\begin{eqnarray}}
\newcommand \eeq{\end{eqnarray}}
 
\begin{document}
 
\title{{\sf Neutrino scattering rates in the presence of hyperons} \\
       {\sf from a Skyrme model in the RPA approximation} 
\author{L. Mornas\footnote{lysiane@pinon.ccu.uniovi.es} \\
{\small {\it Departamento de F{\'\i}sica, Universidad de Oviedo,
Avda Calvo Sotelo 18, 33007 Oviedo, Spain}}}           
}
\maketitle 
 
\begin{abstract}
The contribution of $\Lambda$ hyperons to neutrino scattering rates 
is calculated in the random phase approximation in a model where the 
interaction is described by a Skyrme potential. 
Finite temperature and neutrino trapping are taken into 
account in view of applications to the deleptonization stage of 
protoneutron star cooling. 
The hyperons can remove the problem of ferromagnetic instability 
common to (nearly) all Skyrme parametrizations of the nucleon-nucleon 
interaction. As a consequence, there is not any longer a pole at the 
transition in the neutrino-baryon cross section. However there still 
remains an enhancement in this region. 
In the absence of ferromagnetism the mean free path in $np\Lambda$
matter is reduced compared to its value in $np$ matter as consequence
of the presence of this additional degree of freedom.
At high density the results are very sensitive to the choice of the 
$\Lambda$-$\Lambda$ interaction.
\end{abstract}    

\vskip -0.5cm 
\null\hskip 0.33cm {\small PACS numbers: 26.60.+c, 25.30.Pt, 21.60.Jz}
\vskip -2pt \hskip 0.33cm {\small keywords: neutrinos, proto-neutron stars,
non-relativistic RPA, Skyrme potential, hyperons}

\section{Introduction}

During and shortly after the collapse of a supernova, a copious amount of 
neutrinos is produced. The hope that these neutrinos could permit a 
successful explosion in numerical simulations through the Wilson (delayed) 
mechanism, together with the increase of available computational power, 
has triggered an important effort in designing a new generation of computer 
codes. The Boltzmann equation for neutrino transport can now be solved 
simultaneously with the equations for the hydrodynamical evolution of the 
supernova explosion (see {\it e.g.} \cite{SN-boltz}). 
The differential cross section for the various neutrino production, absorption
 and scattering processes is a key input in the Boltzmann equation. These 
calculations also yield the spectral characteristics of the corresponding 
neutrino burst, which are now well within the reach of modern detectors. 
Although the current picture is that neutrinos alone are not enough to 
prevent shock stall, but 3-D fluid flow with convection is also necessary, 
the solution of the Boltzmann equation still seems to be an unavoidable 
requirement of a supernova explosion simulation. It is therefore necessary 
to have as good as possible a description of the neutrino interaction with 
the matter and of the influence of medium effects, chemical composition 
and temperature on this data.

The neutrino-nucleon cross section and the influence of nuclear correlations
on this parameter in the central protoneutron star have been estimated by
several authors both in relativistic and nonrelativistic models
\cite{IP82}--\cite{SLGZ03}. While it is generally found in relativistic
models that medium effects tend to reduce the scattering rate, the 
findings from nonrelativistic models are less clear cut. The nonrelativistic 
calculations on the other hand have been performed with a greater variety 
of assumptions as to the nature of nuclear correlations taken into 
consideration as the relativistic ones.

A key issue is whether a ferromagnetic instability is excited at high 
density. In this case it would appear as a pole in the neutrino-nucleon 
cross section when the nuclear correlations are calculated in the random 
phase approximation \cite{NHV99,M01}. Such a feature is commonly observed 
when the nuclear interaction is described by a Skyrme force. It has been 
traced to the exchange (Fock) term. A ferromagnetic transition is actually 
also possible in relativistic models \cite{BMNQ95,MT01} treated in the 
Hartree-Fock approximation. A relativistic calculation of neutrino-nucleon
scattering at the RPA level and including exchange terms however has yet 
to be performed. Then again in nonrelativistic models the susceptibility 
has been found to increase and a ferromagnetic state to be energetically
unfavourable in recent calculations where the nuclear interaction is 
extracted from Brueckner-Hartree-Fock calculations \cite{MVB03,SLGZ03}.
 
At later times hyperons begin to appear and affect the luminosities as 
well as the equation of state \cite{PRPLM99}. One can hope to detect 
the tail of this neutrino burst (between 20 and 50 seconds after the 
beginning of the signal), which corresponds to the cooling of the still 
hot but partially deleptonized protoneutron star. In these later stages 
the diffusion equation is thought to be a good approximation, the 
neutrino dynamics then enter in the mean free path.
After $\sim 50$ s, the neutron star has cooled to temperatures of the order 
of 1 MeV and achieved full deleptonization, {\it i.e.} the neutrinos 
are no longer trapped. It then enters in a long cooling phase. The
mean free path is no longer the relevant parameter; instead, the 
available cooling codes need as an input the neutrino emissivity.

The aim of this work is to assess the relevance of the hyperonic degree 
of freedom on the neutrino interaction with neutron star matter with 
a simple description of the interactions, before appealing to the heavier 
machinery of microscopical description through the Brueckner-Hartree-Fock 
calculations or taking into account the relativistic effects. We have 
chosen for this purpose the Skyrme interaction parametrized by Lanskoy 
{\it el al.} \cite{LY97,L98,YBZ88}. This description includes 
the $\Lambda$ hyperons with parameters fitted with the available data on 
hypernuclei. It is also able to give a description of neutron stars in 
agreement with the results of more refined models \cite{M04a}.

Our model is described in section \S \ref{S:formalism} with a presentation 
of the chosen Skyrme parametrizations in \S \ref{S:Skyforces} -- 
\S \ref{S:skyparams} and of the necessary formalism for the calculation 
of the cross section in \S \ref{S:rpacrossec}. Section \S \ref{S:ferro} 
discusses the influence of hyperons on a possible ferromagnetic instability 
and section \S \ref{S:delept} gives an estimate of the evolution of 
the hyperon fraction during the deleptonization of the proto-neutron 
star. We then present results for the vector and axial response 
functions and differential cross section in section \ref{S:respfuncs} 
and for the mean free path in protoneutron star matter in section 
\S \ref{S:mfp}. Section \S \ref{S:concl} summarizes the main results 
and gives a brief discussion of how the model could be improved.

\section{Formalism}
\label{S:formalism}

\subsection{Skyrme forces}
\label{S:Skyforces}

Besides the usual parameterization of the nucleon-nucleon interaction
\beq
V_{NN}(r_1 -r_2) &=&  t_0\, (1+x_0 P_\sigma) \delta(r_1-r_2) 
+ {1 \over 2} t_1\, (1+x_1 P_\sigma) \left[ k'{}^2\  \delta(r_1-r_2) 
+ \delta(r_1-r_2)\ k^2 \right] \nonumber \\
&& + t_2\, (1+x_2 P_\sigma)  k'\, \delta(r_1-r_2)\, k + {1 \over 6} t_3\, 
(1+x_3 P_\sigma)\, \rho_N^\alpha \left({r_1+r_2 \over 2}\right) 
\delta(r_1- r_2) , 
\eeq
we use the following Lambda-nucleon and Lambda-Lambda potentials
as suggested {\it e.g.}  by Lanskoy {\it et al} \cite{LY97,L98}
\beq
V_{N\Lambda} (r_N-r_\Lambda) &=& 
u_0\, (1+y_0 P_\sigma) \delta(r_N-r_\Lambda) + {1 \over 2} u_1\, 
\left[ k'{}^2\  \delta(r_N-r_\Lambda) +  \delta(r_N-r_\Lambda)\ k^2 \right] 
\nonumber \\
&& + u_2\, k'\, \delta(r_N-r_\Lambda)\, k + {3 \over 8} u_3\, (1+y_3 P_\sigma) 
\rho_N^\beta \left({r_N+r_\Lambda \over 2}\right)  \delta(r_N-r_\Lambda) \\
V_{\Lambda\Lambda} (r_1-r_2) &=& \lambda_0 \delta(r_1-r_2) +{1 \over 2}
\lambda_1 \left[ k'{}^2\ \delta(r_1-r_2) + \delta(r_1-r_2)\ k^2 \right] 
\nonumber \\
&& + \lambda_2\, k'\, \delta(r_1- r_2)\, k + \lambda_3\, \rho_\Lambda 
\rho_N^\gamma
\eeq
(We haved dropped in these expressions the spin-orbit terms which will not 
be used in the remainder of this paper).

While this model only includes the $\Lambda$ hyperon, and would seem less 
complete than {\it e.g.} the model of Balberg and Gal \cite{BG97} which 
includes all the hyperons, we fixed our choice on the former because these 
authors provide the two-particle potential. This permitted us to derive 
the Landau parameters for the spin 1 as well as for the spin 0 channels.
The Landau parameters in the spin 1 channel are namely necessary 
for the calculation of the axial structure function in the RPA 
approximation. By contrast, Balberg and Gal \cite{BG97} give 
directly a parametrization of the (unpolarized) energy density 
functional. This parametrization can therefore only be used in the 
mean field approximation. While thermodynamical relationships 
can give access to the Landau parameters in the spin 0 channel,
we have no way of determining those of the spin 1 channel.
Since the axial structure function is known to give the dominant 
contribution to the neutrino cross section, this precludes the
use of the model of Balberg in the RPA approximation.
We will nevertheless use the model of Balberg and Gal at the mean 
field level in order to estimate the error committed by neglecting
other hyperons and in particular the $\Sigma^-$. 

Some {\it a posteriori} justification may be given for a model with 
the $\Lambda$ hyperon only: 
{\it (i)} The parametrization of the N$\Lambda$ and $\Lambda\Lambda$
is relatively well grounded on Brueckner-Hartree-Fock calculations and
duly cross-checked with data on single and double $\Lambda$ hypernuclei.
By contrast, experimental data concerning the $\Sigma$ hyperon is
quite scarce.
{\it (ii)} It is argued from recent analysis of $\Sigma^-$ atoms and the 
absence of a clear signal for the formation of bound $\Sigma$ states 
in ($K^-,\pi$) reactions \cite{MFGJ95,BFG97,Dabrowski}, as well as from 
calculations in quark models \cite{TSHT98,KFFNS00}, that the $\Sigma$ 
single particle potential is probably repulsive in nuclear matter. 
$\beta$-equilibrium calculations performed on the basis of this 
assumption predict that the threshold for $\Sigma$ hyperon formation 
in neutron stars is shifted to very large densities. If such be the case, 
they would concern only a minor fraction of the star or even not appear 
at all (see {\it e.g.} \cite{BG97,SBG00}). 
{\it (iii)} We are still concerned here with investigating qualitative 
effects, namely the influence of a new degree of freedom on the scattering 
rate and the effect of hyperon formation on preventing an eventual 
ferromagnetic transition. 

\subsection{Composition of matter in chemical equilibrium}

In neutron star matter we impose that the conditions for $\beta$-equilibrium 
are fulfilled. We write the equality of the chemical potentials
$\hat\mu = (\mu_n + m_n) -(\mu_p + m_p) = \mu_e -\mu_\nu$, 
$\mu_n + m_n = \mu_\Lambda + m_\Lambda$. We must also impose electric 
charge conservation $n_e = n_p$.
At a given baryonic density the electron, proton and hyperon fractions 
are determined by the solution of these three equations.
The chemical potentials are determined by deriving the Skyrme energy density 
functional with respect to the density of corresponding particle. Their
explicit expression was given in a companion paper \cite{M04a}. 
The electrons are relativistic 
and their chemical potential is given by $\mu_e=\sqrt{k_{Fe}^2 + m_e^2}$.
In protoneutron star matter with trapped neutrinos we have $\mu_\nu = 
(6 \pi^2 n_\nu)^{(2/3)}$, while in colder neutrino-free neutron star matter
$\mu_\nu =0$. The muons were not considered for simplicity, as their 
presence is not expected to lead to qualitative changes in the results.
The threshold density for $\Lambda$ hyperon formation is determined by the 
condition $\mu_\Lambda(thr) = \mu_\Lambda(k_{F\Lambda}=0) = \mu_n + m_n 
- m_\Lambda$. The hyperons form around 2-3 $n_{\rm sat}$ in neutrino-free 
matter, they appear at higher density when neutrinos are trapped in the star.

\subsection{Skyrme parametrizations of the hyperonic sector}
\label{S:skyparams} 

In \cite{M04a} we have examined many combinations of parameters for the
Skyrme forces among those available in the literature in the light of their
suitability for the description of neutron star physics. We settled on a
recent parametrization by Reinhard and Flocard \cite{RF95} in the nucleon
sector SkI3 (or the very similar SkI5) together with the set YBZ6 for the 
nucleon-$\Lambda$ interaction among those recommended by Lanskoy {\it et al.} 
\cite{LY97,YBZ88} as offering one of the best description of hypernuclear 
data. For the $\Lambda$-$\Lambda$ interaction we then considered two
options, either the SLL2 force of Lanskoy \cite{L98} or no interaction.
Lanskoy fitted the binding excess to $\Delta B_{\Lambda\Lambda} = -4.8$ MeV,
{\it i.e.} more atttractive than the presently accepted value of $-1$ MeV. 
The truth should hopefully lie between these two choices.

\begin{center}
\begin{tabular}{|l|ccccccccc|}
\hline
model & $\alpha$ & $t_0$ & $t_1$ & $t_2$ & $t_3$ & $x_0$ & $x_1$ & $x_2$ 
& $x_3$ \\
\hline
SLy10 & 1/6 & 2506.77 & 430.98 & -304.95 & 13826.41 & 1.0398 & -0.6745
      & -1.0  & 1.6833 \\
SkI3  & 1/4 & -1762.88 & 561.608 & -227.09 & 8106.2 & 0.3083 & -1.1722
      & -1.0907 & 1.2926 \\ 
SV    & 1  & -1248.3 & 970.6 & 107.2 & 0. & -0.17 & 0. & 0. & 1. \\   
\hline
\end{tabular}
\medskip
\newline
{\bf Table I}: Skyrme parameters for the NN interaction 
\cite{RF95,Sly,BFvGQ75}
\end{center}

\begin{center}
\begin{tabular}{|l|ccccccc|}
\hline
model & $\beta$ & $u_0$ & $u_1$ & $u_2$ & $u_3$ & $y_0$ & $y_3$ \\
\hline
LY-I & 1/3 & -476. & 42. & 23. & 1514.1 & -0.0452 & -0.280 \\
YBZ-6  & 1 & -372.2 & 100.4 & 79.60 & 2000. & -0.107 & 0. \\
\hline
\end{tabular}
\hfil
\begin{tabular}{|l||cc|}
\hline
model & $\lambda_0$ & $\lambda_1$ \\
\hline
SLL2  &  -437.7     &     240.    \\
no LL &     0       &       0     \\
\hline
\end{tabular}
\medskip
\newline
{\bf Table II}: Skyrme parameters for the N$\Lambda$ and 
$\Lambda\Lambda$ interactions \cite{LY97,L98,YBZ88}
\end{center}

All expected properties for the neutron star structure were then recovered.
Very importantly, the ferromagnetic transition common to (nearly) all Skyrme
models is avoided when hyperons are present (see \S \ref{S:ferro}). 

Among the main results obtained with the SkI3+YBZ6+SLL2 parametrization, 
let us mention:
\par\indent {\bf -} The equation of state is softened by the presence of the
hyperon degrees of freeedom. As a consequence, the maximum mass 
(2.26 $M_\odot$) for a $npe$ star is reduced to 1.64 $M_\odot$ for a star 
with hyperons. 
\par\indent {\bf -} A hot star with trapped neutrinos was observed to be
metastable in the mass range [1.64 - 1.88] $M_\odot$, as noticed in several
earlier studies, see {\it e.g.} \cite{PBPELK97}. A protoneutron star
in this mass range will eventually collapse to a black hole as it undergoes 
deleptonization.
\par\indent {\bf -} The equation of state remains causal in the whole star
when hyperons are present.
\par\indent {\bf -} The $\Lambda$ hyperons appear at a threshold density
$2.08\ n_{\rm sat}$ in cold neutrino-free matter. In protoneutron star 
matter with a lepton fraction $Y_L=0.4$, the formation of hyperons 
is delayed until $2.85\ n_{\rm sat}$.
\par\indent {\bf -} The asymmetry energy $a_A = (1/2) \partial(E/A)/\partial
\beta^2|_{\beta \rightarrow 0}$ increases rapidly with density with the
SkI3 parametrization, allowing for high proton fractions. 
\par\indent {\bf -} There exists a non negligible hyperon fraction in
the core of the protoneutron star before deleptonization is fully
completed (see section \S \ref{S:delept}). The tail of the supernova 
neutrino burst which corresponds to this stage could therefore be affected 
by the presence of hyperons.

Another suitable set was the Skyrme Lyon SLy10 \cite{Sly} together with 
the LYI parametrization of the N-$\Lambda$ force also favored by Lanskoy. 
The asymmetry energy increases with density but much slower than with 
the SkI3 force; as a consequence, the proton fraction remains rather low. 
With hyperons the maximum mass of the neutron star was a lower but still 
acceptable 1.425 $M_\odot$. The set SLy10 was preferred over the more 
widely used SLy4 and SLy7 sets because it was easier to keep it clear of 
the ferromagnetic pole. In other respects the properties of SLy10 are 
very similar to that of the Sly4 and Sly7 sets.

Finally we singled out the older SV force \cite{BFvGQ75}. This set is rather
atypical as it has $t_3=0$, its effective mass at saturation is also
lower ($m_* = 0.38$) than the commonly accepted values. Although its 
adequacy for reproducing the properties of nuclei and nuclear matter
is somewhat less satisfactory than was the case for SkI3 or SLy10,
it has the unique feature of not presenting a ferromagnetic transition in
nuclear matter nor in $npe$ matter in $\beta$-equilibrium which makes 
it very valuable for comparison purposes. Its general characteristics
(stiffness and maximum neutron star mass, proton content) are quite 
similar to those of the SkI3 set.

We reproduce below the relevant entries of the tables published in 
\cite{M04a} 

\begin{center}
\begin{tabular}{|c||c|c||c|}
\hline
 Model    & $n_{\rm thr}^{\Lambda}$ & $Y_p$ & $n_{\rm ferro}$  \\
\hline
SLy10+LY-I+SLL2 & 2.719 & 0.041 & 6.183 \\
SkI3+YBZ6+SLL2  & 2.076 & 0.144   & 9.194 \\
SV+YBZ6+SLL2    & 1.951 & 0.136  & 9.093 \\
\hline
\end{tabular}
\vskip 0.3cm
\begin{minipage}{12cm}
\begin{flushleft}
{\bf Table III}: Thresholds for hyperon formation and for ferromagnetism 
in $np\Lambda e$ matter in $\beta$ equilibrium
\end{flushleft}
\end{minipage}
\end{center}     

\begin{center}
\begin{tabular}{|c||c||c|c||c|c|c|}
\hline
Model  & $n_B(c_s^2=1)$ & $n_c (1.4\ M_\odot)$ & $R (1.4\ M_\odot)$
& $n_{\rm max}$ & $M_{\rm max}$ & $R(M_{\rm max})$ \\
\hline
SLy10, $npe$ matter & 7.308  & 3.60  & 11.05 &  7.69 & 1.99  & 9.52 \\ 
SLy10+LY-I+SLL2     & 16.032 & 9.38  &  9.0  & 12.38 & 1.425 & 8.11 \\
SkI3, $npe$ matter  & 6.343  & 2.27  & 13.21 &  6.12 & 2.263 & 11.16 \\ 
SkI3+YBZ6, no LL    & causal & 2.36  & 13.20 & 4.66  & 1.655 & 12.52 \\
SkI3+YBZ6+SLL2      & 13.077 & 2.39  & 13.19 & 6.79  & 1.642 & 11.02 \\
SV, $npe$ matter    & 4.983  & 2.10  & 13.46 & 5.44  & 2.426 & 11.54 \\
SV+YBZ6, no LL      & causal & 2.19  & 13.40 & 4.44  & 1.662 & 12.75 \\
SV+YBZ6+SLL2        & 13.082 & 2.22  & 13.41 & 6.5   & 1.639 & 11.24 \\
\hline
\end{tabular}
\vskip 0.3cm
\begin{minipage}{12cm}
\begin{flushleft}
{\bf Table IV}: Conditions for causality and neutron star properties
\end{flushleft}
\end{minipage}
\end{center}

\subsection{Neutrino-baryon scattering and absorption rates}
\label{S:rpacrossec}

In the non relativistic approximation the differential scattering cross 
section for the process $ \nu(p_1) + B (p_2) \rightarrow \nu (p_3) + 
B(p_4)$ is taken to be (see {\it e.g.} \cite{IP82,BRT04,NHV99,M01}):
\beq
{1 \over V} {d \sigma \over d \omega d \Omega} = {G_F^2 \over 8 \pi^3}
(E_\nu')^2  [ 1-f(E_\nu') ] \left[ (1 + \cos \theta) {\cal S}^{(0)}(\omega,k)
+ (3 - \cos \theta) {\cal S}^{(1)}(\omega,k) \right]
\label{eq:diffcross}
\eeq
In this expression, $\omega$ and $k$ are the transferred energy and momentum
$p_1^\mu -p_3^\mu = (\omega, \vec k)$, $E_\nu'=E_\nu-\omega$ is the neutrino
energy after the collision and $\theta$ is the scattering angle
$\vec p_1.\vec p_3 = |p_1|\, |p_3|\, \cos\theta$. 
The structure functions ${\cal S}^{(S)}$ in spin channel $S$ are defined by
the expectation value of the density and spin density fluctuations in
baryonic matter. In the mean field approximation they both reduce to 
\beq
{\cal S}_0 = 2\, \int {d^3\, p_2 \over (2 \pi)^3}\, {d^3\, p_4 
\over (2 \pi)^3} f(p_2) (2 \pi)^4 \delta^4 (p_1+p_2-p_3-p_4)
\left(1-f(p_4)\right)
\eeq
They are related to the imaginary part of the reducible polarizations by:
\beq
{\cal S}^{(S)}(\omega, k) = -2 {1 \over 
1 -e^{-\beta(\omega +\hat \mu)}}\ {\cal I}{\rm m}\, \Pi^{(S)}(\omega, k)
\eeq 
This definition differs by a factor 1/$\pi$ from that of 
\cite{NHV99,M01,GRNvGS92} but agrees with that of \cite{BS98,BRT04} so that
the elastic limit be ${\cal S}_0(\omega,k) \rightarrow 2 \pi n_B 
\delta(\omega)$ 

In the non relativistic limit, the spin 0 and spin 1 responses correspond 
to the vector and axial couplings respectively. In the random phase 
approximation the vector and axial structure functions obey uncoupled 
Dyson equations:
\beq
\Pi_{V/A}^{RPA} = \Pi_0 + \Pi_{V/A}^{RPA}\ V_{(S=0/1)}\ \Pi_0,
\label{Dyson}
\eeq
where 
\beq
\Pi_0 = {2 \over (2 \pi)^3} \int d^3 p\ \left[ {f(E_p) -f(E_{p+q}) 
\over q_0 + E_p -E_{p+q} +i \epsilon} \right] 
\eeq
and $V_{(S=0/1)}$ the interaction potential in the particle-hole channel.

\vskip 0.2cm
In our case, the Dyson equation has a 3x3 matrix structure.
The RPA polarization is obtained by inverting (\ref{Dyson}) with
\beq
\Pi_0=\left[ \matrix{ \Pi_n^0 & 0 & 0 \cr
                      0 & \Pi_p^0 & 0 \cr
                      0 & 0 & \Pi_\Lambda^0 \cr} \right].
\eeq
\noindent
We have for the real part of the polarizations
\beq
{\cal R}{\rm e}\ \Pi_0^i(\omega,k) &=& {m_i^* \over 2 \pi^2 k}\ \int_0^\infty
p dp \ \ln \left[{ (k^2 -2 k p)^2 -(2 m_i^* \omega)^2 \over
(k^2 +2 k p)^2 -(2 m_i^* \omega)^2} \right]\ f(E_i)  
\label{eq:RePi} \\
{\rm with} && f(E_i) = \left\{ 1 +\exp[ \Bigl( {p^2 \over 2 m_i^*} 
-\tilde\mu_i \Bigr)/T]+1 \right\}^{-1} \quad {\rm and} \quad 
\tilde \mu_i =\mu_i -{\cal U}_i \nonumber
\eeq
\noindent
The expressions for the effective masses and chemical potential
were given in the Appendix of \cite{M04a}.
At $T=0$, ${\cal R}{\rm e}\ \Pi_0^i$ reduces to the Linhardt function 
in agreement with {\it e.g.} \cite{GRNvGS92}
\beq
{\cal R}{\rm e}\ \Pi_0^i(\omega,k) & \stackrel{T\rightarrow 0}{\longrightarrow}
   & -{m_i^* p_{Fi} \over 2 \pi^2}\ [1 + \phi_i(x_+) + \phi_i(x_-)]\,  \\
   && \phi_i(x)={p_{Fi} \over 2 k}\ [1-x]\ \ln \left|{1+x \over 1-x} \right| 
   \ , \quad x_\pm = {k \over 2 p_{Fi}} \pm {\omega m_i^* \over k p_{Fi} }
\nonumber
\eeq
The imaginary part is always analytical
\beq
{\cal I}{\rm m}\ \Pi_0^i(\omega,k) = -{(m_i^*)^2 T \over 2 \pi k}\ 
\ln  \left\{ {1 + \exp\left[-\Bigl(\displaystyle{p_-^2\over 2 m_i^*} 
              - \tilde\mu_i \Bigr)/T \right] \over 
              1 + \exp\left[-\Bigl(\displaystyle{p_+^2\over 2 m_i^*} 
              - \tilde\mu_i \Bigr)/T \right] } \right\} \, 
              \quad {\rm with}\quad p_\pm = p_{Fi} x_\pm 
\label{eq:ImPi}
\eeq

We will extract the potential in the Landau Fermi Liquid approximation
from the Skyrme parameterization. Our potential has no imaginary part 
in this approximation. We will be working in the simplest monopolar 
($l=0$) Landau approximation; our potential has therefore no angular 
or momentum dependence. For a discussion of the full RPA with momentum 
dependence we refer the reader to \cite{NHV99,M01,HNP97,BVA95}. 

The potential in the vector isovector channel for neutrino scattering 
through the neutral current process is given in our model in terms of 
the Landau parameters $f_{ij}$
\beq
V_{(S=0)} = \left[ \matrix{ f_{nn} & f_{np} & f_{n\Lambda} \cr
                         f_{pn} & f_{pp} & f_{p\Lambda} \cr
                         f_{\Lambda n} & f_{\Lambda p} & f_{\Lambda\Lambda} 
                         \cr } \right] 
\label{eq:phpot}
\eeq
For the axial channel, one should replace all $f_{ij}$ by the $g_{ij}$ 
in order to obtain the relevant potential $V_{(S=1)}$. The Landau 
parameters can be extracted from the  Skyrme potential energy by applying a
double functional differentiation with respect to occupation numbers
\cite{GRNvGS92,HNP97}. Their explicit expression when hyperons are present
was given in \cite{M04a}.

After performing the inversion, we obtain for example in the axial ($S=1$)
channel
\beq 
\Pi_{A\, nn}^{\rm {\small RPA}} &=& {1 \over {\rm Det} [I - V_{(S=1)}\,
\Pi_0 ]}\ \left[ (1-g_{pp} \Pi_p^0) (1-g_{\Lambda\Lambda} \Pi_\Lambda^0) 
- g_{p\Lambda}^2  \Pi_p^0 \Pi_\Lambda^0 \right] \Pi_n^0 \\
\Pi_{A\, np}^{\rm {\small RPA}} &=& {1 \over {\rm Det} 
[I - V_{(S=1)}\, \Pi_0]}\ \left[ (1-g_{\Lambda\Lambda} \Pi_\Lambda^0) g_{pn}
+ g_{p\Lambda} g_{n\Lambda} \Pi_\Lambda^0 \right]   \Pi_p^0 \Pi_n^0 \\
\Pi_{A\, n\Lambda}^{\rm {\small RPA}} &=& {1 \over {\rm Det} 
[I - V_{(S=1)}\, \Pi_0]}\ \left[  (1-g_{pp} \Pi_p^0) g_{n\Lambda}
+ g_{np} g_{p\Lambda} \Pi_n^0 \right]  \Pi_n^0 \Pi_\Lambda^0 
\eeq
with similar expressions for the vector channel $(S=0)$. These equations
reduce to Eqs. (26-32) of Reddy {\it et al.} \cite{RPLP99} when no hyperons
are present.

Let us note at this stage that if the determinant appearing in the
denominator is strongly reduced or vanishes, this will give rise
to a corresponding peak in the structure function and in the 
neutrino cross section. This is indeed what happens in the axial 
channel for previous calculations \cite{NHV99,M01} in pure neutron 
matter and $npe$ matter in $\beta$ equilibrium (see also Fig. 16 of 
\cite{RPLP99}): The condition  
\beq
D_A = {\rm Det} [I - V_{(S=1)}\, \Pi_0] =0
\label{eq:denoA}
\eeq
is fulfilled at some critical density, signalling the transition
to a ferromagnetic state. This is discussed further in next section.

Finally the structure functions appearing in equation (\ref{eq:diffcross})
are obtained from 
\beq
S_{V/A} = \left( \matrix{ c_{V/A}^p & c_{V/A}^n & c_{V/A}^\Lambda \cr} 
          \right)\, . \,
          \left( \matrix{S_{V/A}^{pp} & S_{V/A}^{pn} & S_{V/A}^{p\Lambda} \cr
          S_{V/A}^{np} & S_{V/A}^{nn} & S_{V/A}^{n\Lambda} \cr
          S_{V/A}^{\Lambda p} & S_{V/A}^{\Lambda n} & S_{V/A}^{\Lambda\Lambda}
          \cr} \right) \, . \, 
          \left( \matrix{ c_{V/A}^p \cr c_{V/A}^n \cr c_{V/A}^\Lambda \cr} 
          \right)
\eeq 
with 
\beq
c_V^p &=& {1 \over 2} (1 -4 \sin^2\theta_W) = 0.04\ , 
\quad c_V^n = -{1 \over 2} \ , \quad c_V^{\Lambda} = -{1 \over 2} 
\nonumber \\
c_A^p &=& {1 \over 2} (D+F) = {1.26 \over 2}\ , \quad c_A^n = 
-{1 \over 2} (D+F) = -{1.26 \over 2}\ , \quad c_A^{\Lambda } =  -{1 \over 2} 
(F+D/3) = -0.365 \nonumber \\
\eeq

\subsection{Poles in the dispersion relations}
\label{S:ferro}

Previous studies on the neutrino mean free path in neutron matter
\cite{NHV99} or $npe^-$ matter in $\beta$ equilibrium \cite{M01}
found that a pole appears in the calculation of the axial structure 
function above a certain critical density. This feature is typical
of Skyrme parameterization of nuclear interactions and is related to 
a transition to a ferromagnetic state. 

In \cite{M04a} we defined the magnetic susceptibilities
$\chi_{ij}$ where $i,j \in \{ n,p,\Lambda \}$.
The inverse susceptibilities are proportional to the second 
derivatives of the energy density functional with respect to the 
polarizations:
\beq
{\mu_i \mu_j \over \chi_{ij} } &=& {2 \rho \over \rho_i \rho_j} \Delta_{ij} \\
\Delta_{ij} &=& {1 \over 2} {\partial^2 ({\cal E}/\rho) \over \partial
s_i \partial s_j} \\
s_i &=& {\rho_{i\uparrow} -  \rho_{i\downarrow} \over
       \rho_{i\uparrow} +  \rho_{i\downarrow}}
\eeq
On the other hand it can be shown that the $\Delta_{ij}$ are related 
to the Landau parameters $g_0^{ij}$  through
\beq
\Delta_{ij} &=& {2 \rho \over \rho_i \rho_j} {1 \over \sqrt{ N_0^i N_0^j}} 
G_0^{ij} \quad {\rm if} \ i \ne j \nonumber \\
\Delta_{ii} &=& {2 \rho \over \rho_i^2} {1 \over N_0^i} 
(1 + G_0^{ii} ) \nonumber \\
G_0^{ij} &=& \sqrt{N_0^i N_0^j} g_0^{ij} \quad, \qquad  
N_0^i= {m_i^* k_{Fi} \over \pi^2 \hbar^2}
\eeq
A criterion for the appearance of the ferromagnetic phase is that 
the determinant of the inverse susceptibility matrix vanishes,
or equivalently, in terms of the Landau parameters:
\beq
{\rm Det} \left( \matrix{ (1 + G_0^{nn}) & G_0^{np} & G_0^{n\Lambda} \cr
                    G_0^{pn} & (1 + G_0^{pp}) & G_0^{p\Lambda} \cr
    G_0^{\Lambda n} & G_0^{p\Lambda} & (1 + G_0^{\Lambda\Lambda}) \cr} 
\right) =0
\label{eq:ferro}
\eeq
In the limit where the temperature $T$ and the energy transfer $\omega$ 
go to zero, the real part of the Linhardt functions tend to the limiting 
value
\beq
{\cal R}{\rm e}\, \Pi_0^i(\omega,k)\ \stackrel{ {}^{(T\rightarrow 0,\, 
     \omega \rightarrow 0,\, \omega/k = {\rm cst})} } {\longrightarrow} 
     \ - N_0^i
\eeq
If we further assume that the imaginary part of the polarizations 
vanish (which, even though it turns out to be an excellent approximation, 
is not strictly the case) we would then find that the condition that 
a pole appears in the dressed axial polarization ${\rm Det} 
[I - V_{(S=1)} \Pi_0]$ [{\it cf.} Eq. (\ref{eq:denoA})] is identical to the 
criterion for the appearance of a ferromagnetic phase defined in this 
section. In fact the finite imaginary parts provide for some Landau damping 
of the pole. The pole will also be somewhat shifted and smoothed by finite 
temperature, but this hardly changes our conclusions in practice for 
the temperatures relevant for protoneutron stars (up to 30 MeV).
 
In \cite{M04a} we plotted the criterion (\ref{eq:ferro}) as a function
of density. It was shown that when the threshold for hyperon formation was
lower than the critical density for the ferromagnetic transition in
$npe$ matter, the hyperons were able to stabilize the system so that
the pole could be avoided. An sample of these results is displayed
in Fig. 1.  

\begin{figure}[htb]
\mbox{%
\parbox{7.5cm}{\epsfig{file=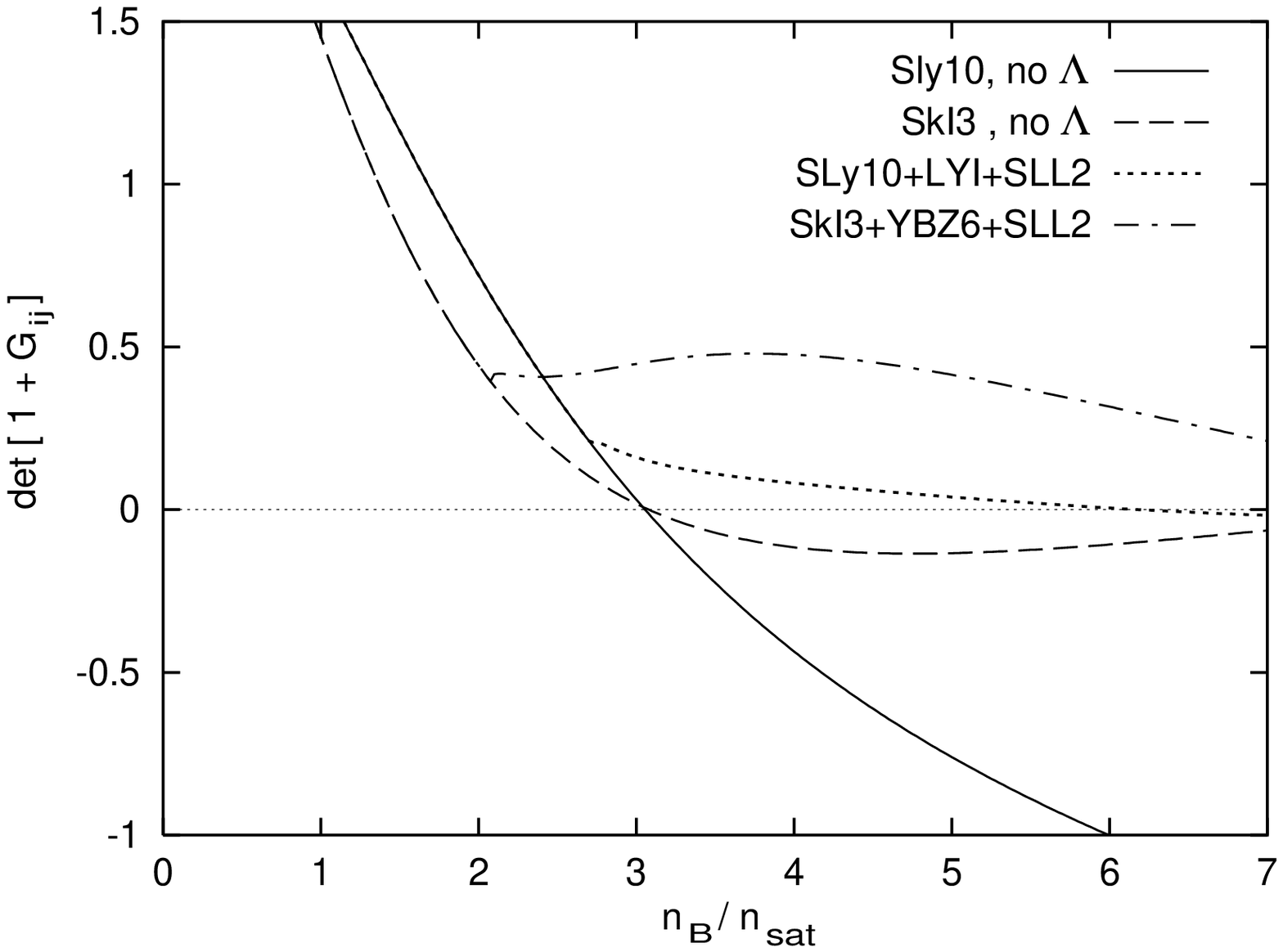,width=7.5cm}}
\parbox{1cm}{\phantom{aaaaa}}
\parbox{6.5cm}{{\bf Fig. 1} -- Criterion for the transition to 
a ferromagnetic state.}
}
\end{figure} 

A similar study can be performed for the vector response function.
The denominator $D_V = {\rm Det} [I - V_{(S=0)}\, \Pi_0] =0$ reduces 
at $T=0$, $\omega \rightarrow 0$ to the criterion for instability under 
density fluctuations equivalent to Eq. (\ref{eq:ferro}) by replacing 
the $G_{ij}$ by the corresponding $F_{ij}$.  At low density, no hyperons 
are present, and the criterion reduces to that for spinodal instability 
in asymmetric nuclear matter (see {\it e.g.} \cite{Baran}) 
$(1+F_0^{nn})(1+F_0^{pp})-(F_0^{np})^2=0$.
The pole disappears at temperatures $\sim$ 15 MeV, but the enhancement
persists at higher temperatures.

\section{Evolution of the hyperonic content in a deleptonizing 
proto neutron star}
\label{S:delept}

As the protoneutron star cools and the trapped neutrinos are released, 
the beta equilibrium becomes more favorable to the production of hyperons.
If a sizable number of hyperons appear before the deleptonization is
completed, the mean free path of the neutrinos will be affected and 
could influence the characteristics of the tail of the neutrino burst 
detected in supernova explosions. The determination of the evolution of 
the hyperonic content would require to solve self consistently the 
equations for thermodynamical conditions, neutrino mean free path,
diffusion of neutrinos and cooling along the lines of reference
\cite{PRPLM99}. 

In this section we take the results from Fig. 17 of Pons {\it et al.}
\cite{PRPLM99} as representative of the evolution the temperature and 
leptonic content. With this input we calculate the structure of the 
neutron star at each time step. While this procedure is not consistent,
it still permits to estimate whether hyperons will be relevant at this 
stage.

We chose the evolution curves correponding to model GM3 with hyperons
(curved labeled GM3npH in \cite{PRPLM99}). The temperature $T$ and total 
lepton number $Y_L=Y_e+Y_\nu$ are fixed to values extracted from this 
figure (see table below). The central density was also taken according 
to the results of \cite{PRPLM99} but was rescaled for each parameter 
set in order to match with a 1.6 $M_\odot$ neutron star at the end of 
deleptonization. It was checked that the equation of state and chemical 
composition at the end of deleptonization obtained by setting $Y_L=0.09$ 
coincides with that obtained when setting $Y_\nu=0$.

\vskip 0.2cm

\begin{tabular}{|l|ccccccccccc|}
\hline
$t$ [s]   &   0  &   5    &   10  &  15  &  20  &  25  &  30  &  35  &  40  & 
             45  &  50 \\
$T$ [MeV] & 17.3 &  25.0  &  31.7 & 36.5 & 37.5 & 36.5 & 33.6 & 29.8 & 26.0 & 
            22.1 & 18.3 \\
$Y_L$& 0.345 & 0.315 & 0.283 & 0.256 & 0.239 & 0.222 & 0.202 & 0.178 & 
       0.145 & 0.119 & 0.09 \\
\hline
\end{tabular}

\begin{figure}[htb]
\mbox{%
\parbox{8cm}{\epsfig{file= 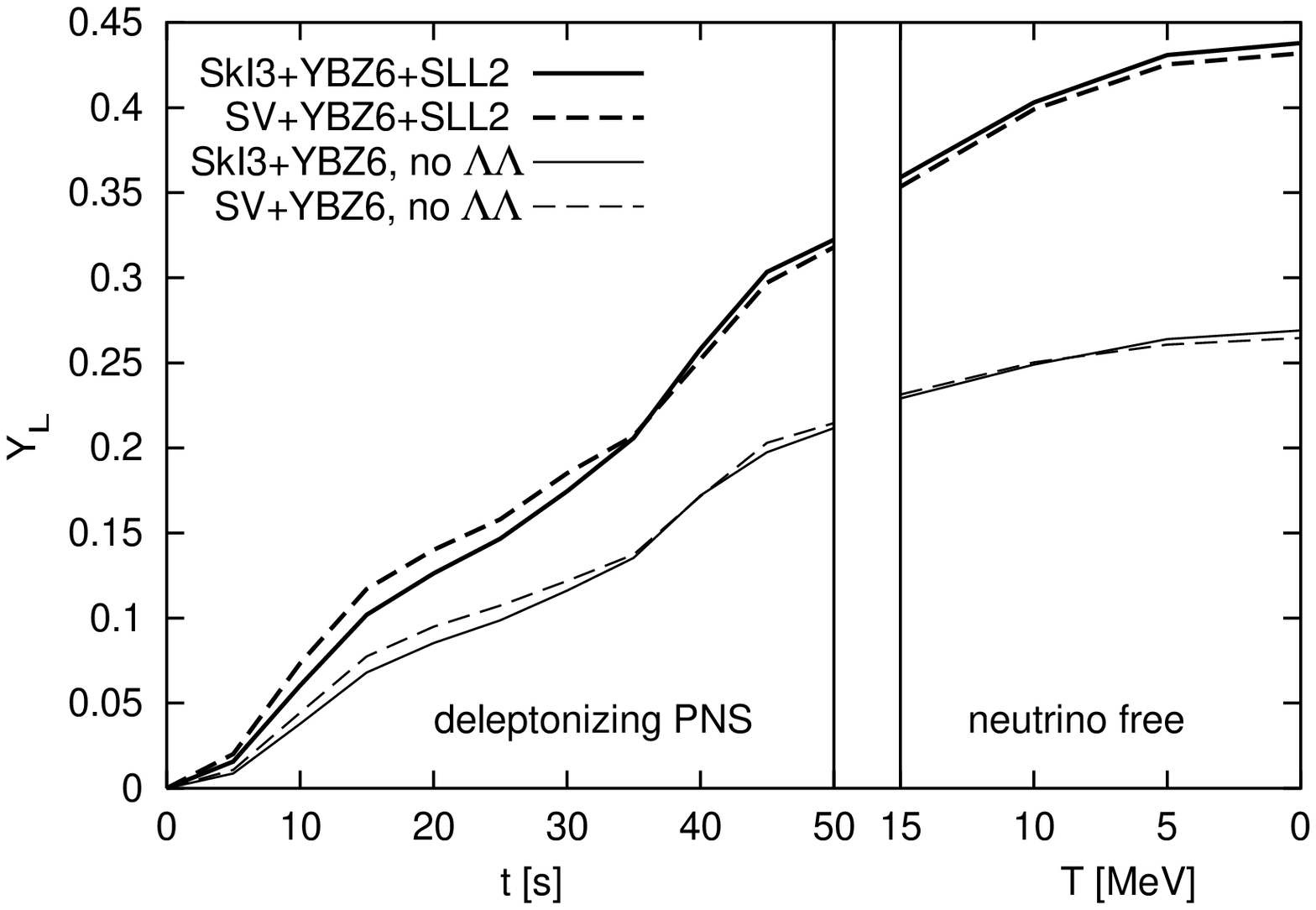,width=8cm}}
\parbox{1cm}{\phantom{aaaa}}
\parbox{7cm}{{\bf Fig. 2} --  Evolution of the hyperonic content as 
a function of time}}
\end{figure} 

\begin{figure}[htb]
\begin{minipage}{8cm}
\mbox{%
\parbox{8cm}{\epsfig{file=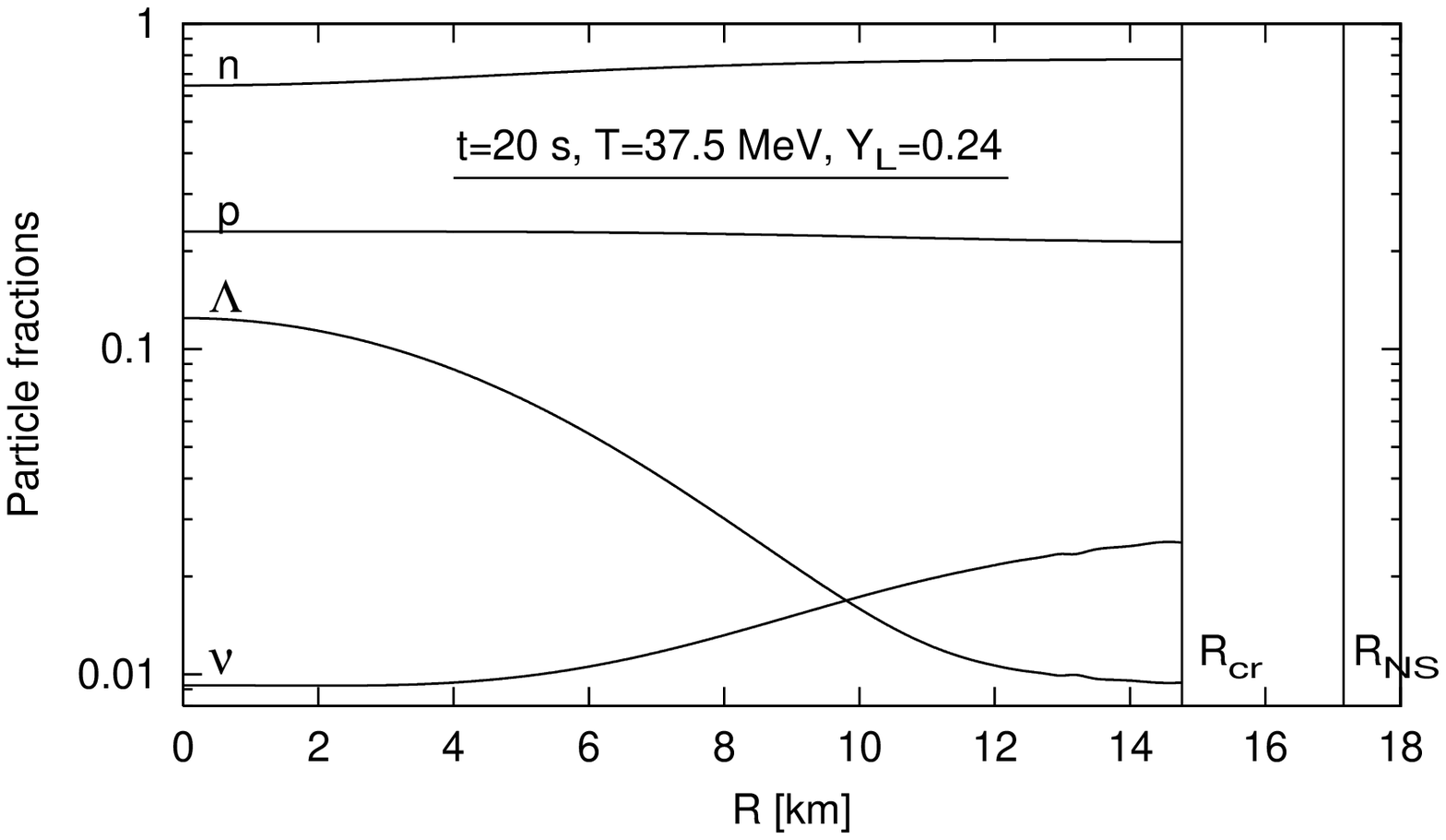,width=8cm}}
}
\mbox{%
\parbox{8cm}{\epsfig{file=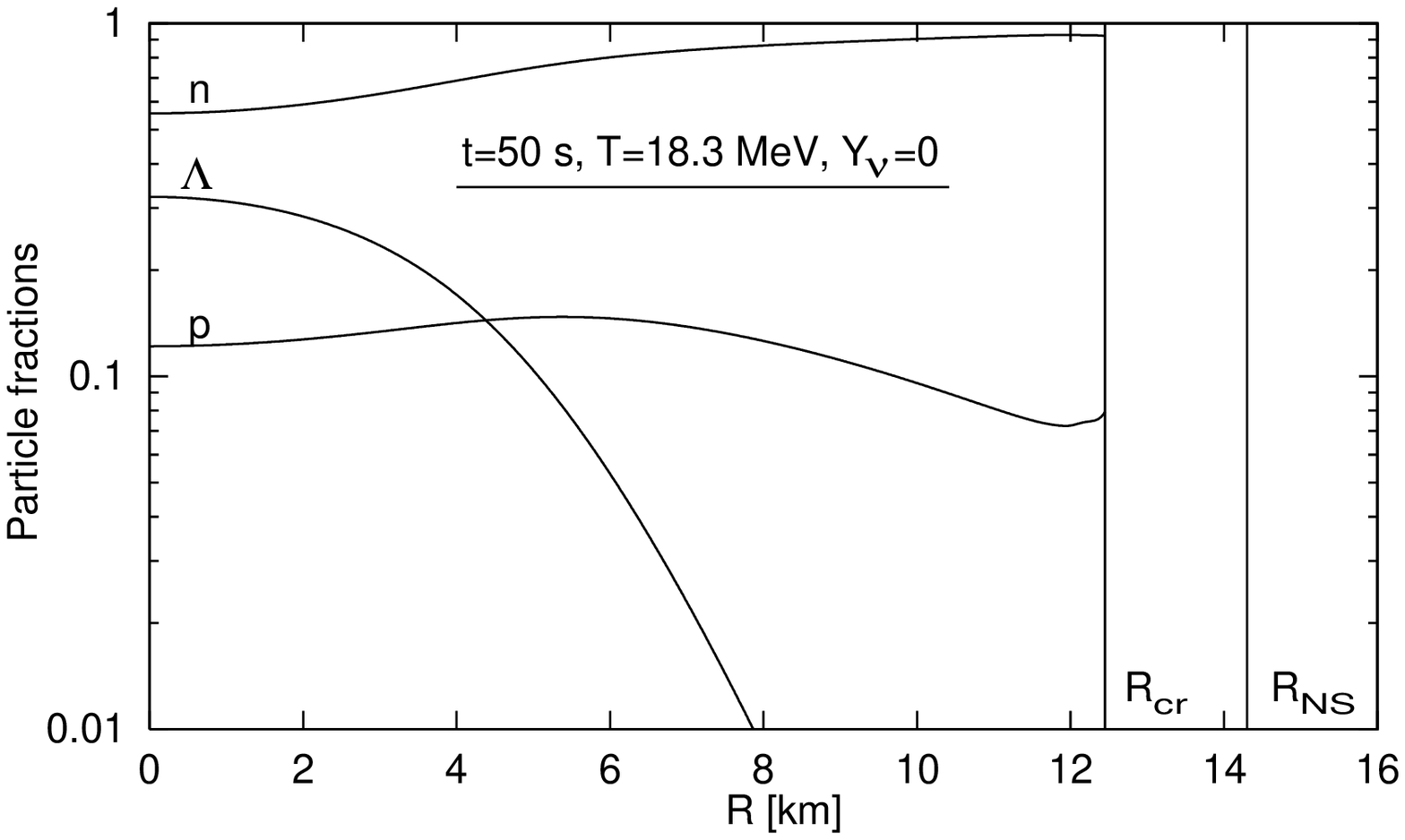,width=8cm}}
}
\mbox{%
\parbox{8cm}{\epsfig{file=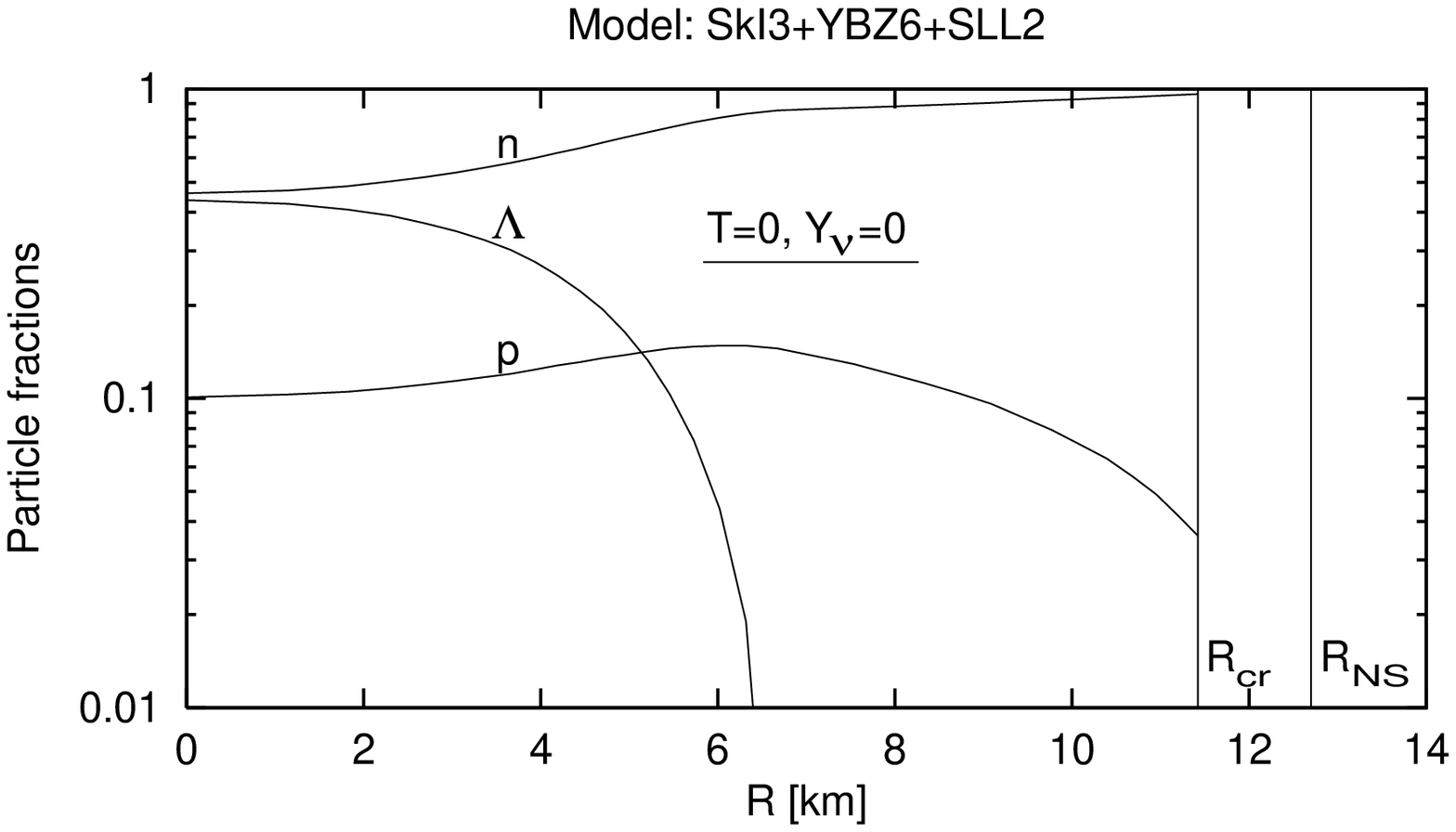,width=8cm}}
}
\end{minipage} 
\hskip 0.2cm
\begin{minipage}{8cm}
\mbox{%
\parbox{8cm}{{\bf Fig. 3} --  Particle fractions as
a function of the neutron star radius at three stages of the 
deleptonization. }}
\end{minipage} 
\end{figure} 

\vskip 0.2cm

After neutrinos have left the neutron star, it continues to cool from 
$T=18.3$ MeV to T=0. We perfomed four more calculations at T=0, 5, 10 and 15
MeV to follow the hyperonic content in this later stage. The result is 
displayed in Figures 2 and 3 for the parameter sets which allow a stable
neutron star with a mass 1.6 $M_\odot$, namely SkI3 or SV in the 
nucleon sector, YBZ6 for the hyperon-nucleon interaction and either
no Lambda-Lambda interaction or the SLL2 parametrization of Lanskoy
{\it et al.}

In Fig. 2 the hyperonic fraction at the center of the star is displayed 
as a function of time. We can see that a sizable fraction of hyperons is 
present as early as 10 seconds after the collapse. The result are very 
similar whether we use the SkI3 or SV force. On the other hand we again 
note a strong dependence on the characteristics of the $\Lambda$-$\Lambda$
interaction.

Fig. 3 shows the density profile of the neutron star at three stages
of the deleptonization ($t$=20s, $t$=50s and $t \rightarrow \infty$) 
for the parameter set SkI3+YBZ6+SLL2. As the star cools, its radius
shrinks and the hyperons gather at its center.

\section{Neutrino scattering rates}

\subsection{Structure functions}
\label{S:respfuncs} 

\par\noindent$\underline{\mbox{\sf a) Axial response}}$
\vskip 0.2cm

Let us first discuss the behaviour of the contributions to the 
axial structure functions from neutrons ($S_A^{(nn)}$), protons 
($S_A^{(pp)}$) and $\Lambda$ hyperons ($S_A^{(\Lambda\Lambda)}$). 

\begin{figure}[htb]
\mbox{%
\parbox{7.5cm}{\epsfig{file=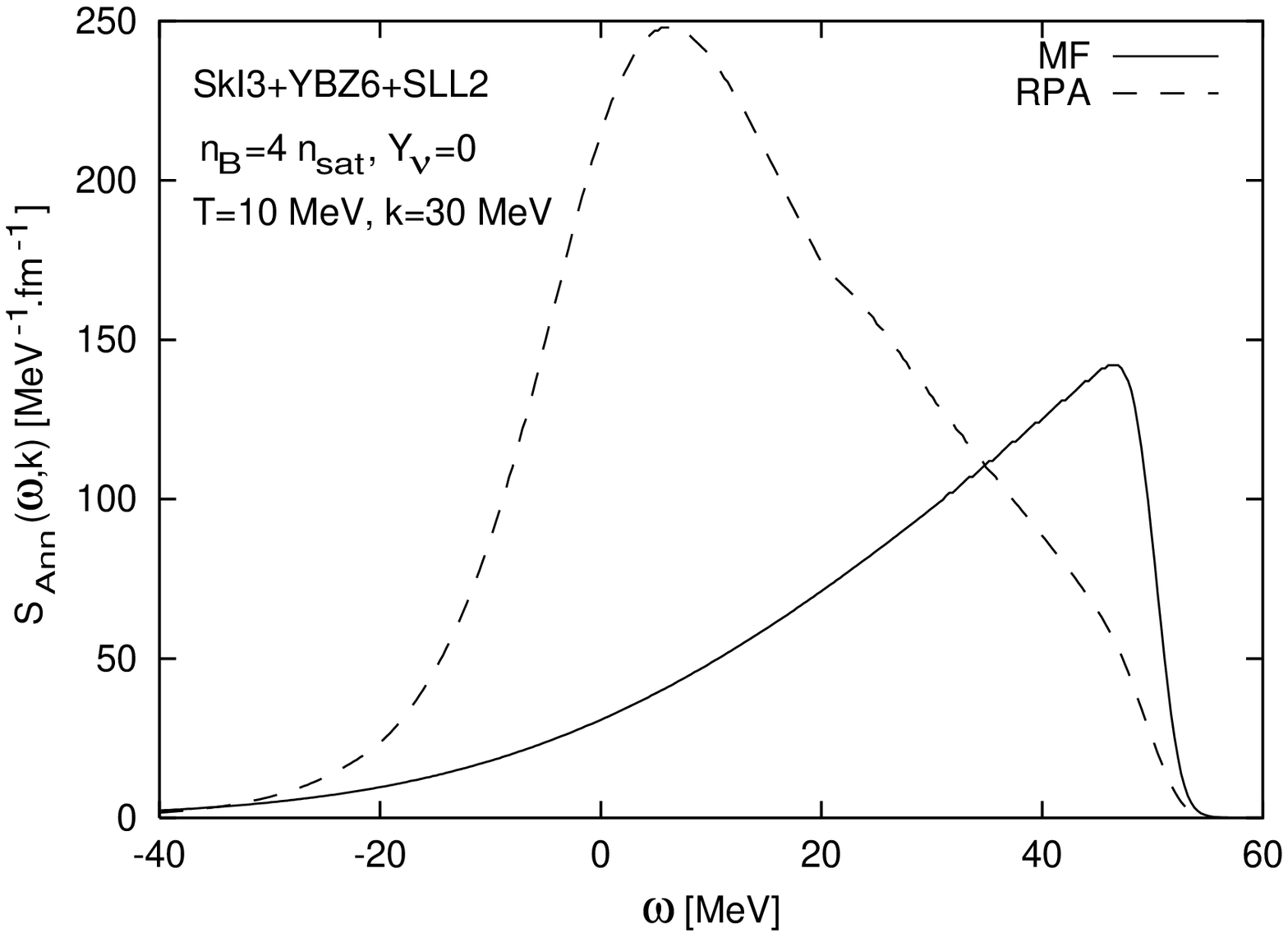,width=7.5cm}}
\parbox{7.5cm}{\epsfig{file=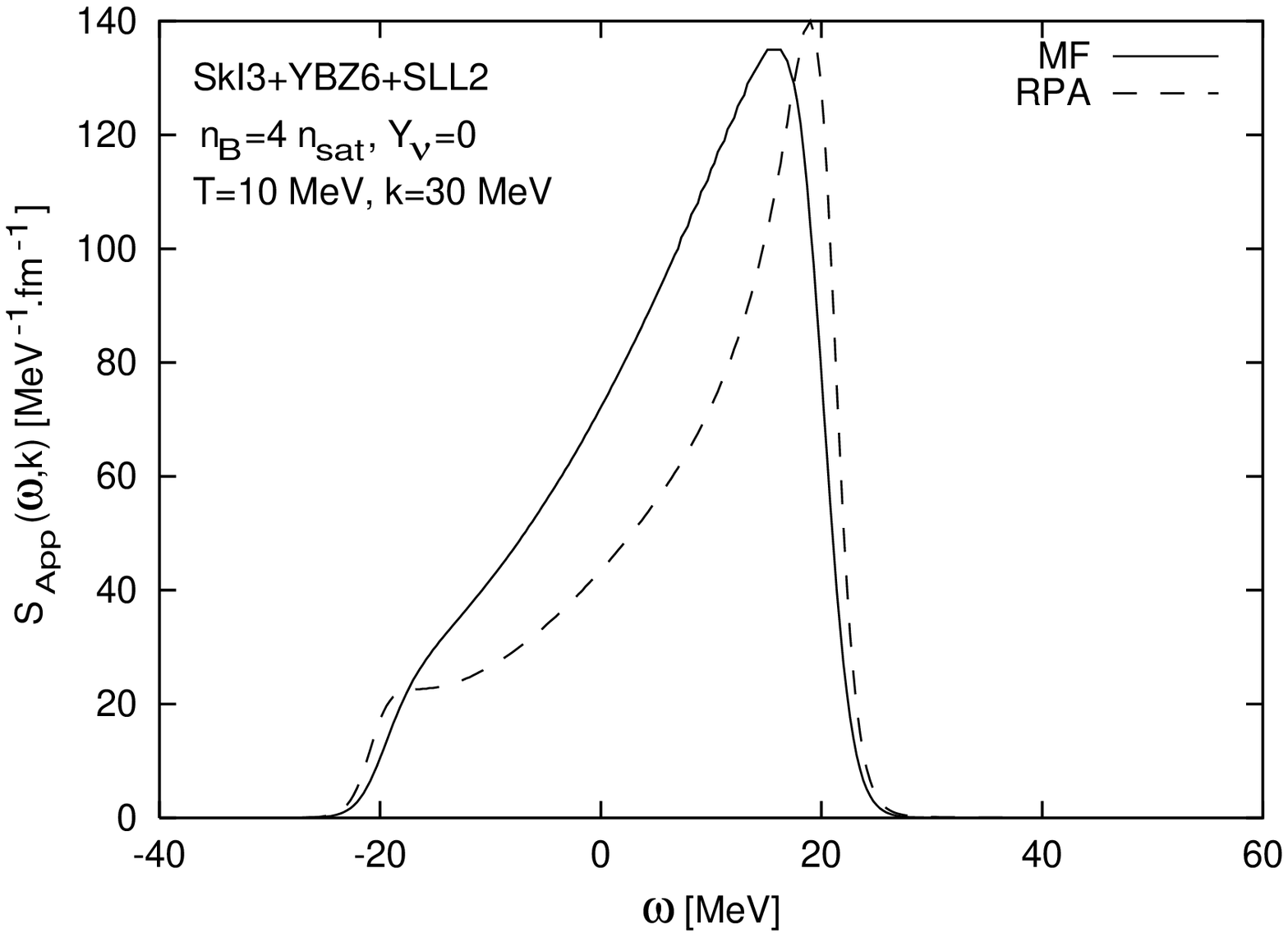,width=7.5cm}}
}
\mbox{%
\parbox{7.5cm}{\epsfig{file=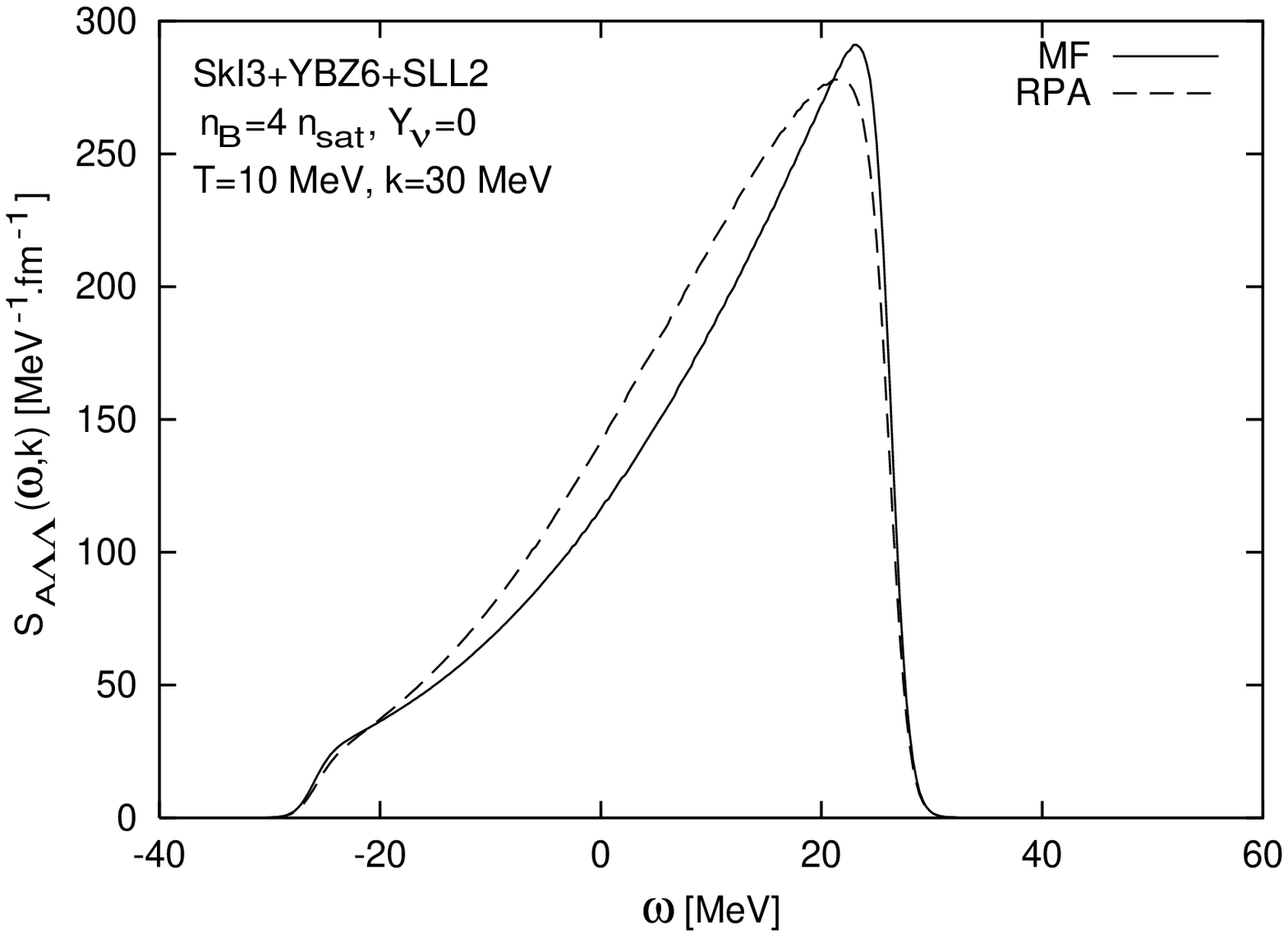,width=7.5cm}}
\parbox{1cm}{\phantom{aaaaa}}
\parbox{6.5cm}{{\bf Fig. 4} -- Axial structure functions in
$npe\Lambda$ matter in beta equilibrium -- Mean field and RPA 
approximations}
}
\end{figure} 

As already obtained by other authors \cite{NHV99,M01}, the axial 
structure function for neutrons in $npe$ matter displays a strong 
enhancement at $\omega=0$, going over to a pole at the critical density 
$n_{\rm ferro}^{\beta-npe}$. We have seen in \S \ref{S:ferro}, 
the criterion for ferromagnetic instability is equal to the condition 
that the denominator of the dressed polarizations vanish at $\omega=0$.
This is a mechanical instability, meaning that an infinitesimal energy 
fluctuation $\omega=\varepsilon$ can trigger the phase transition. This 
should not be confused with the phenomenon of spin zero sound analogous 
to the zero sound mode observed in the vector reponse function, where 
a peak appears at a finite value $\omega=\omega_{ZS}$. In the latter 
case the response function displays a resonance corresponding to the 
excitation of a collective motion of the system at the corresponding 
velocity $v_{ZS}$.

Fig. 4 compares the axial structure functions obtained in the mean field 
and random phase approximations. The system is $npe\Lambda$ matter in 
$\beta$ equilibrium at a density $n_B=4\ n_{\rm sat}$, temperature 
$T=10$ MeV and  we fixed the transferred momentum at $k=30$ MeV. 
The interactions are described by the choice of parameters 
SkI3 + YBZ6 + SLL2.  While the RPA correction is only moderate
in the $S_A^{(pp)}$ and $S_A^{(\Lambda\Lambda)}$ contributions, 
we can clearly see the enhancement due to the vicinity of the ferromagnetic
criterion in $S_A^{(nn)}$. On the other hand, no spin zero sound 
peak is observed at a finite value of $\omega$.

\begin{figure}[htb]
\mbox{%
\parbox{7.5cm}{\epsfig{file=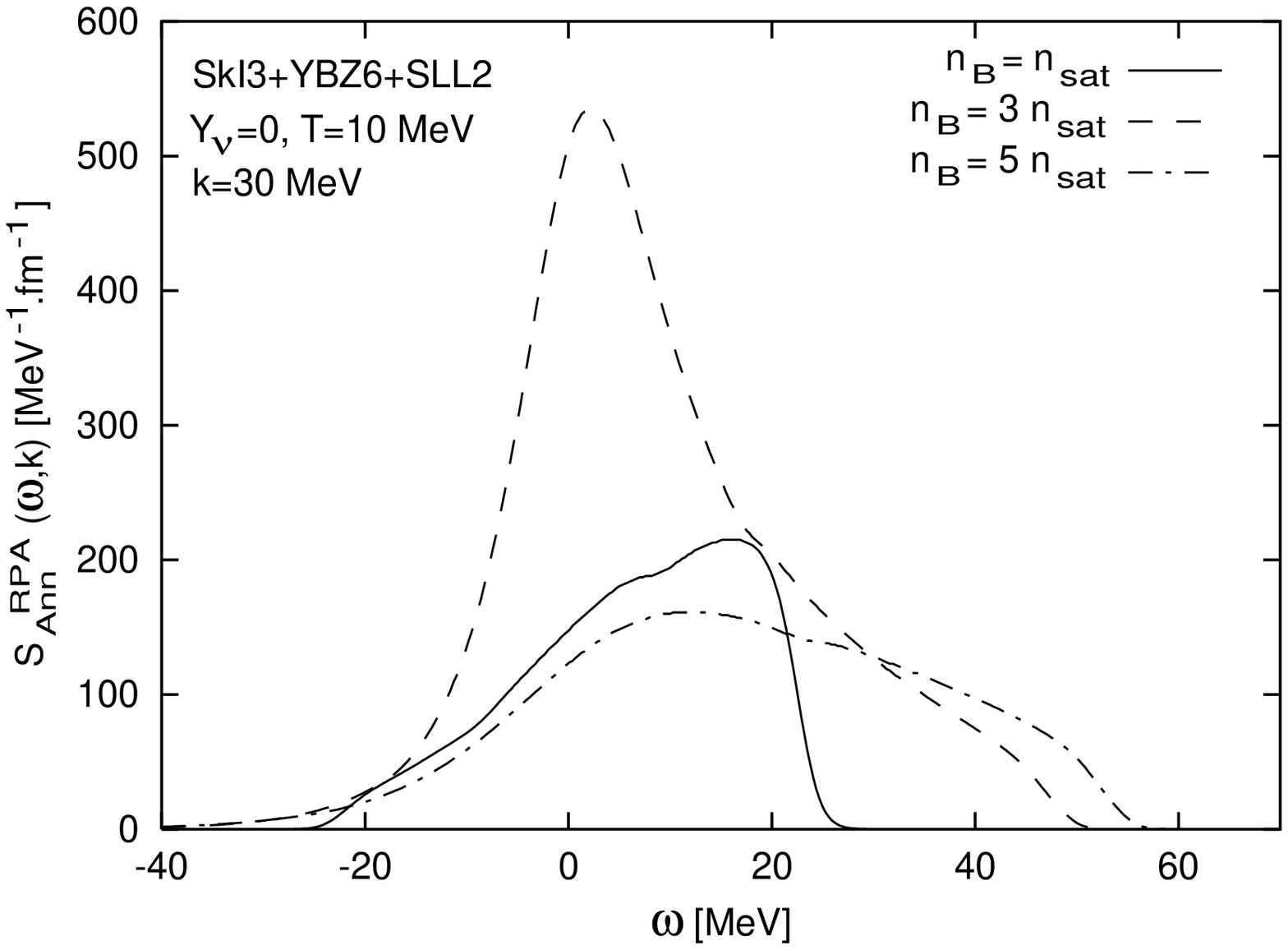,width=7.5cm}}
\parbox{7.5cm}{\epsfig{file=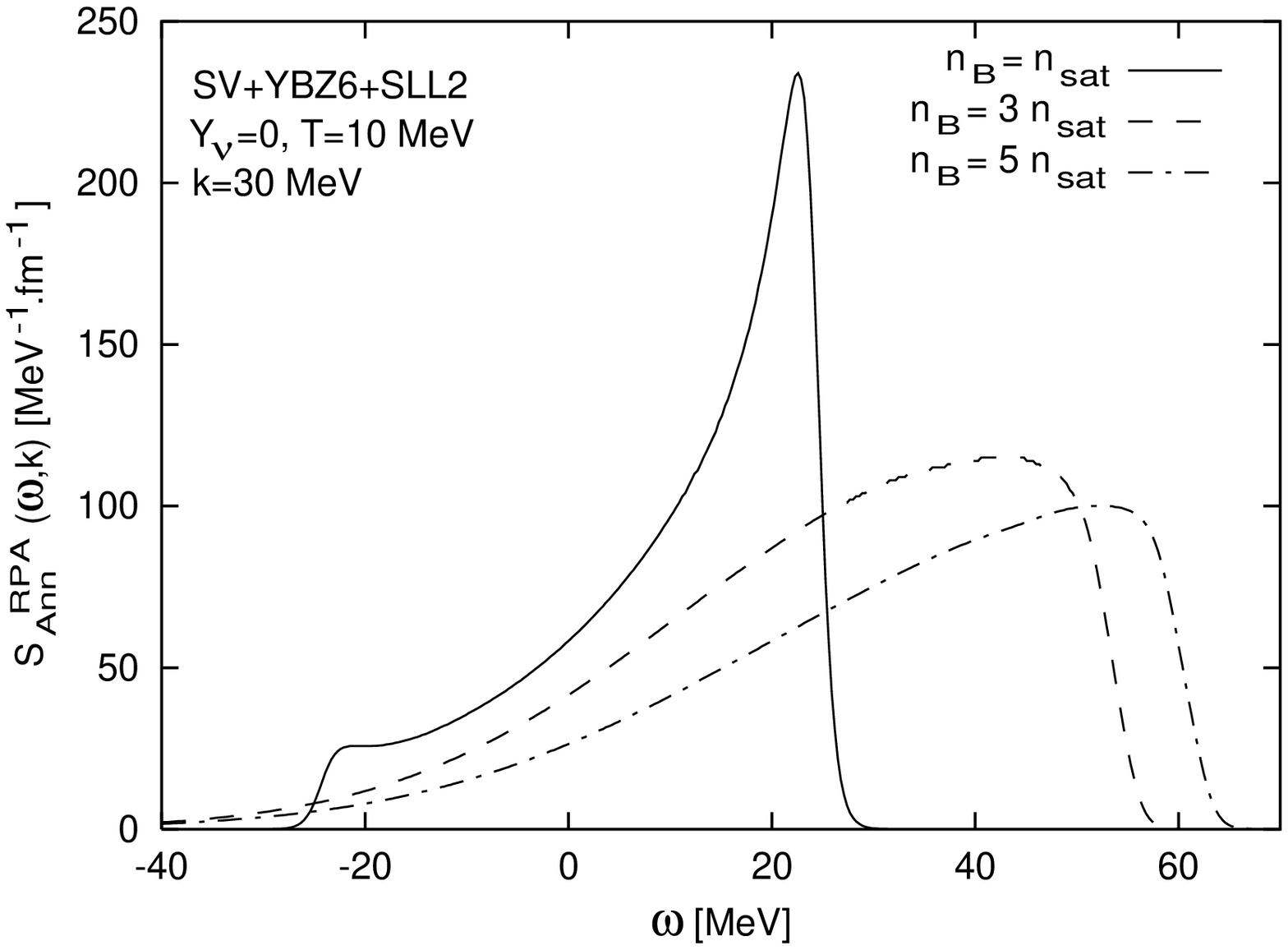,width=7.5cm}}
}
\mbox{%
\parbox{7.5cm}{\epsfig{file=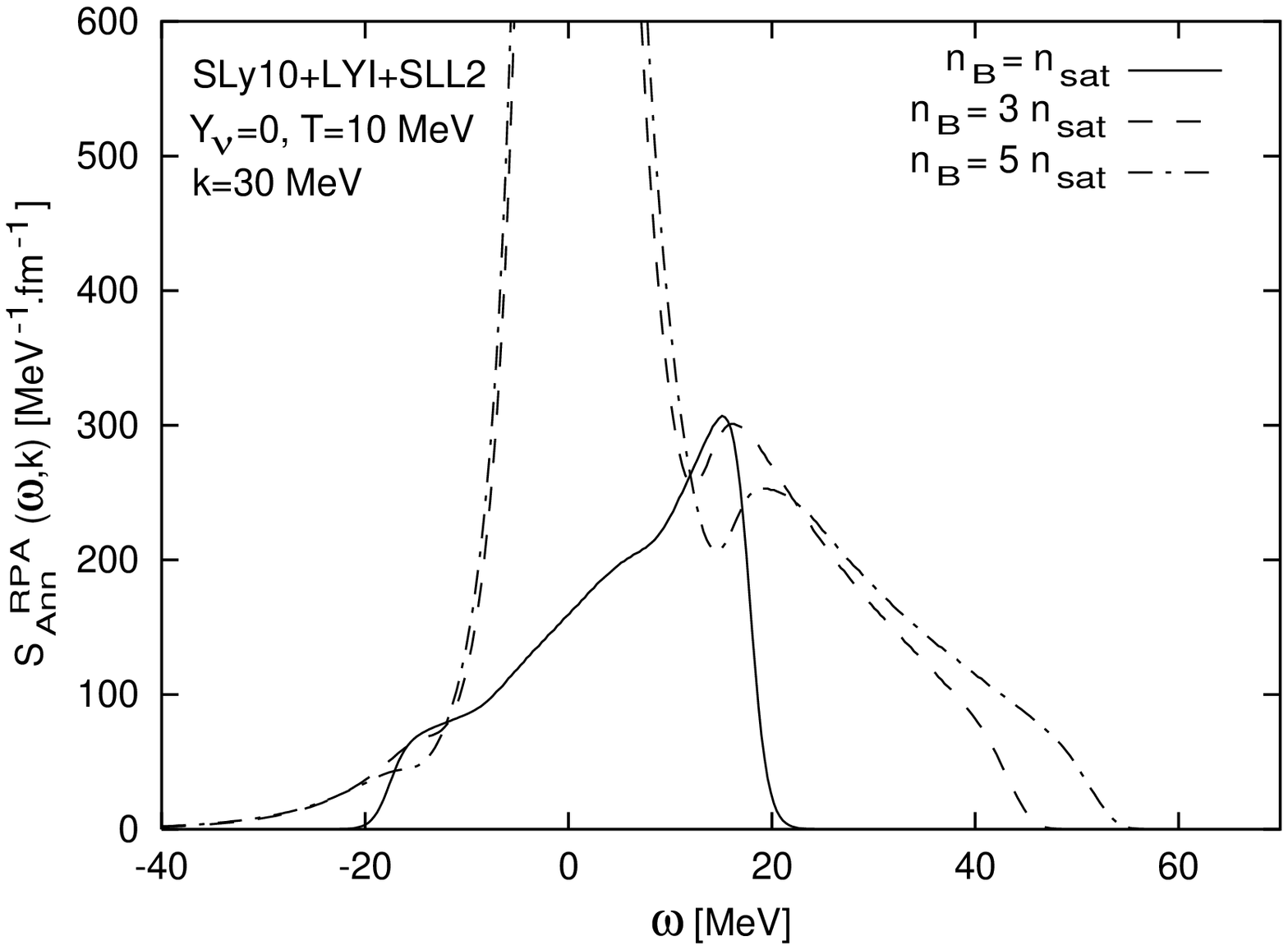,width=7.5cm}}
\parbox{1cm}{\phantom{aaaaa}}
\parbox{6.5cm}{{\bf Fig. 5} -- Axial structure functions in
$npe\Lambda$ matter in beta equilibrium in the RPA approximation
-- Comparison of three Skyrme parametrizations}
}
\end{figure} 

Fig. 5 compares the $S_A^{(nn)}$ as a function of density for three 
different Skyrme parametrizations: SLy10+LYI+SLL2, SkI3+YBZ6+SLL2 and
SV+YBZ6+SLL2. At saturation the three parametrizations give very similar
results. At three time saturation density on the other hand the quantity
$D_A(\omega=0)={\rm det}(1+G_{ij})$ is approaching the zero axis for
the SkI3 and SLy10 models (see Fig. 1) and the $S_A^{(nn)}$ response is
strongly enhanced. The SV+YBZ6+SLL2 parametrization which is free of 
ferromagnetism accordingly does not give rise to any enhancement at 
$\omega=0$.

\begin{figure}[htb]
\mbox{%
\parbox{7.5cm}{\epsfig{file=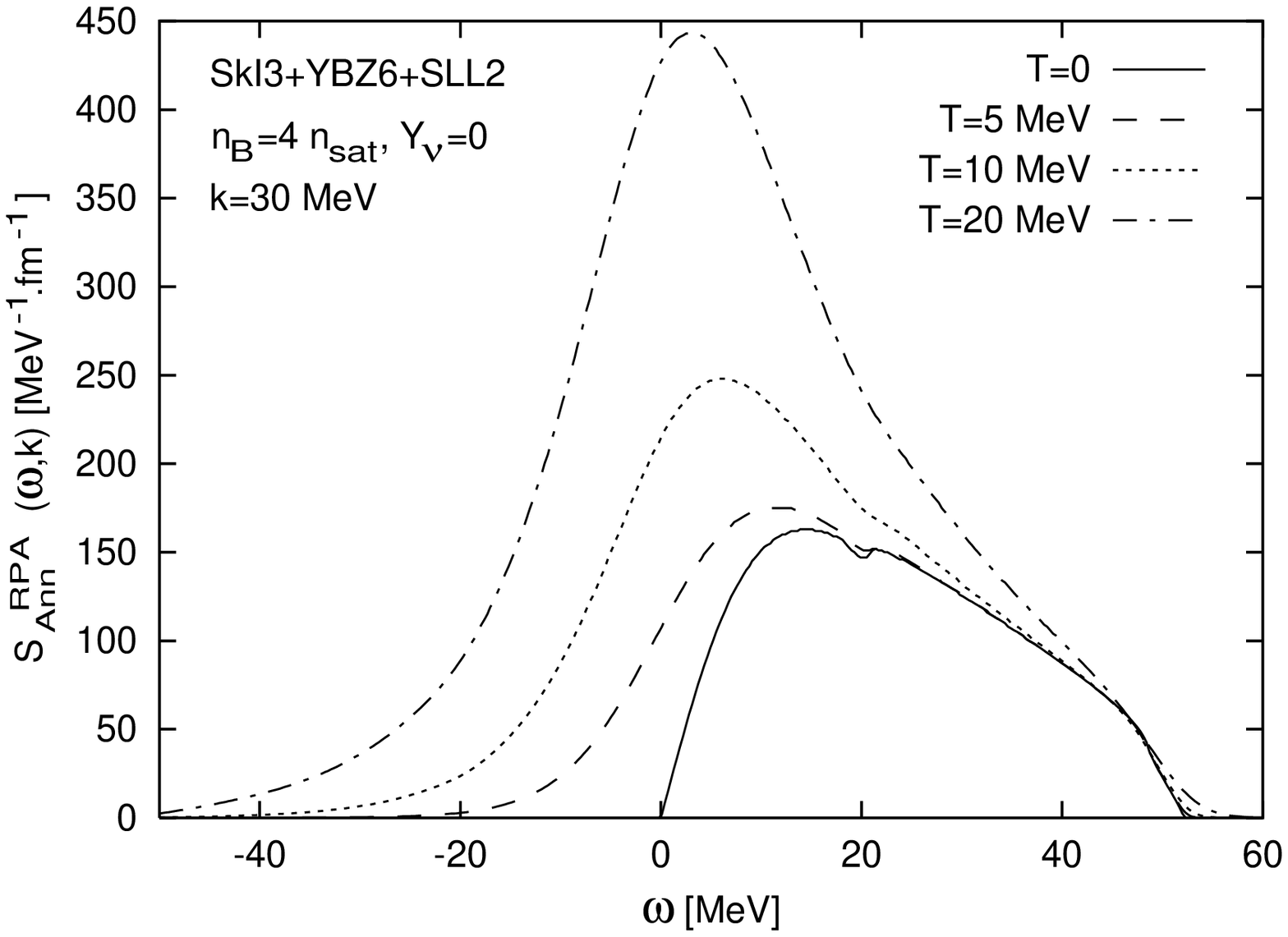,width=7.5cm}}
\parbox{1cm}{\phantom{aaaaa}}
\parbox{7.5cm}{\epsfig{file=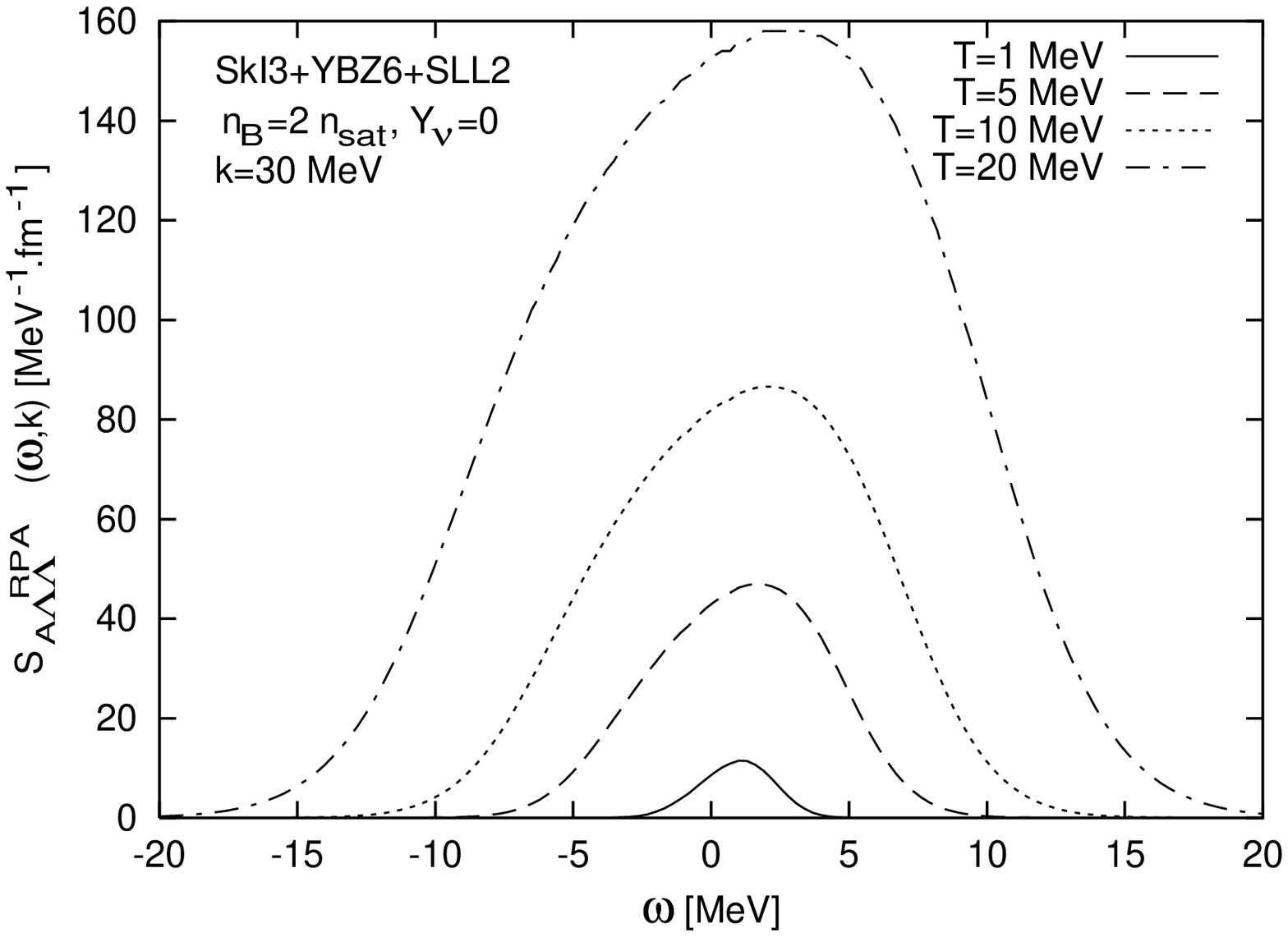,width=7.5cm}}
}
\vskip 0.2cm
\mbox{%
\parbox{16cm}{{\bf Fig. 6} -- Axial structure functions in
$npe\Lambda$ matter in beta equilibrium in the RPA approximation
-- Temperature dependence}
}
\end{figure} 

$S_A^{(pp)}$ follows the usual trend of having its strength shifted 
to higher $\omega$ when density is increased. This is also observed 
for  $S_A^{(nn)}$ in the SV+YBZ6+SLL2 parametrization when the effect 
is not washed out by the enhancement at $\omega=0$. At $T=0$
$S_A^{(\Lambda\Lambda)}$ is only non zero above the threshold 
for hyperon production and then qualitatively behaves as $S_A^{(pp)}$. 
At high density its contribution is of the same order of magnitude as 
that of the neutrons.

Fig. 6 illustrates the temperature dependence of the axial structure
functions. On the left panel, $S_A^{(nn)}$ is represented at $n_B
=4\ n_{\rm sat}$ and $k$=30 MeV. As the temperature is increased,
the response function is enhanced and processes in which the neutrino 
gain energy in the collision ($\omega <0$) gradually open.
On the right panel, the hyperon structure function is displayed at 
$n_B=2\ n_{\rm sat}$, {\it i.e.} shortly below the threshold density
$n_{\rm thr}^\Lambda=2.08\ n_{\rm sat}$. While the structure function 
vanishes as expected at zero temperature, its contribution increases
as temperature rises and rapidly reaches the same order of magnitude
as scattering on nucleons: Even though there are only traces of 
thermally created hyperons in this regime, they contribute actively 
to the scattering as they are not yet impeded by the Pauli blocking.

\vskip 0.2cm
\par\noindent$\underline{\mbox{\sf b) Vector response}}$
\vskip 0.2cm

In Fig. 7 we show the vector structure function for the neutron. At low 
density we find when plotting $D_V(\omega,k)={\rm Det} [I - V_{(S=0)}
\Pi_0 ]$ that this function comes near to the zero axis and eventually 
crosses it. This can be explained as the onset of the spinodal instability, 
when homogeneous matter is not the stablest state anymore, as is the case 
at the inner edge of the crust of the neutron star (see {\it e.g.} 
\cite{NScrust}). As the density is increased, this instability disappears 
but instead dips form in $D_V$ for a finite value of the energy transfer. 
When the real part of $D_V$ vanishes a zero sound mode develops. This occurs 
at approximately $3\ \rho_{\rm sat}$ depending on the equation of state. 
At moderate densities, the imaginary part is finite and provides for Landau 
damping of the mode. This is seen on the vector structure function as a 
peak with some width. At higher density the zero sound mode is unscreened 
and appears as a delta function outside the vector strength distribution.

\begin{figure}[htb]
\mbox{%
\parbox{7.5cm}{\epsfig{file=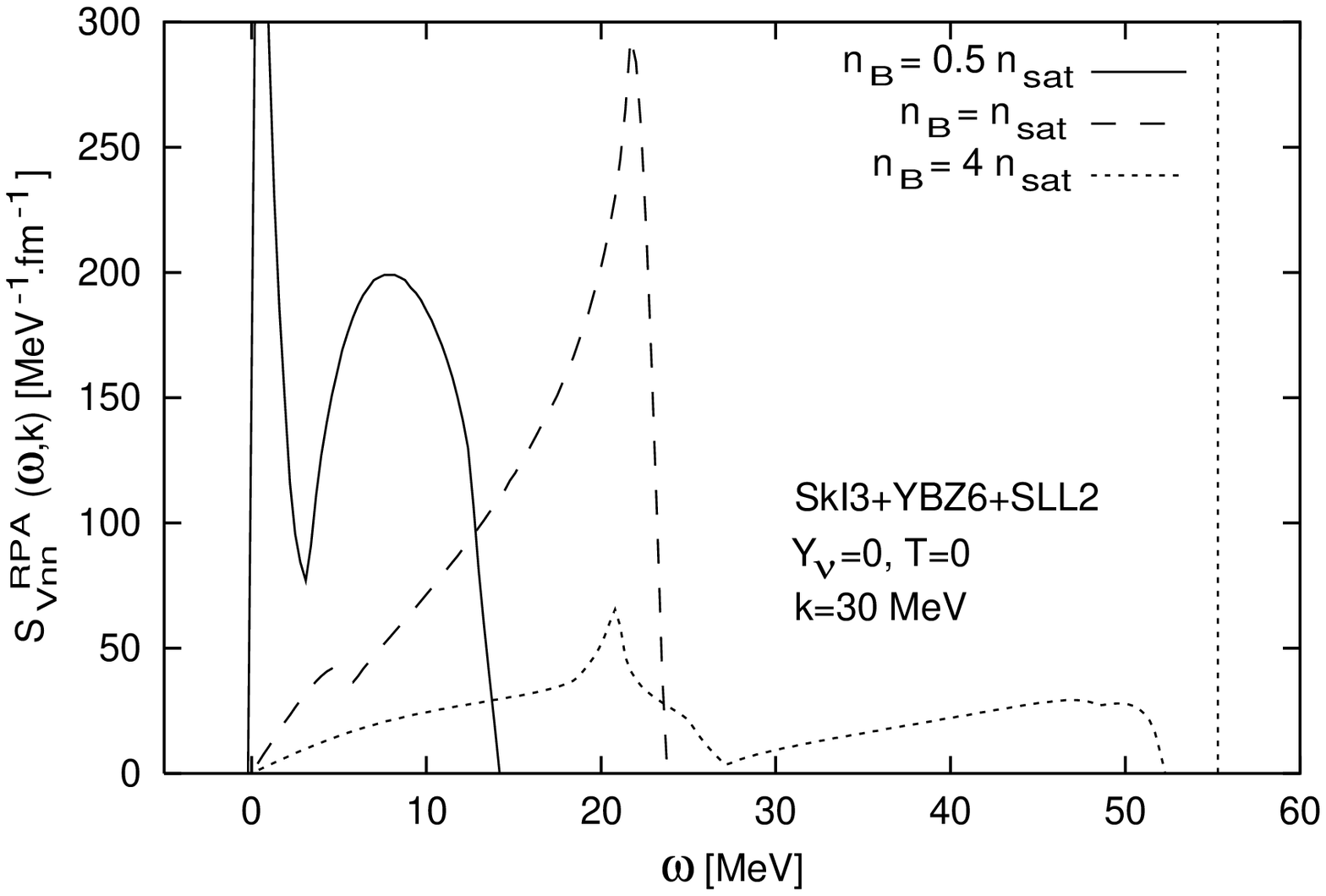,width=7.5cm}}
\parbox{1cm}{\phantom{aaaaa}}
\parbox{7.5cm}{\epsfig{file=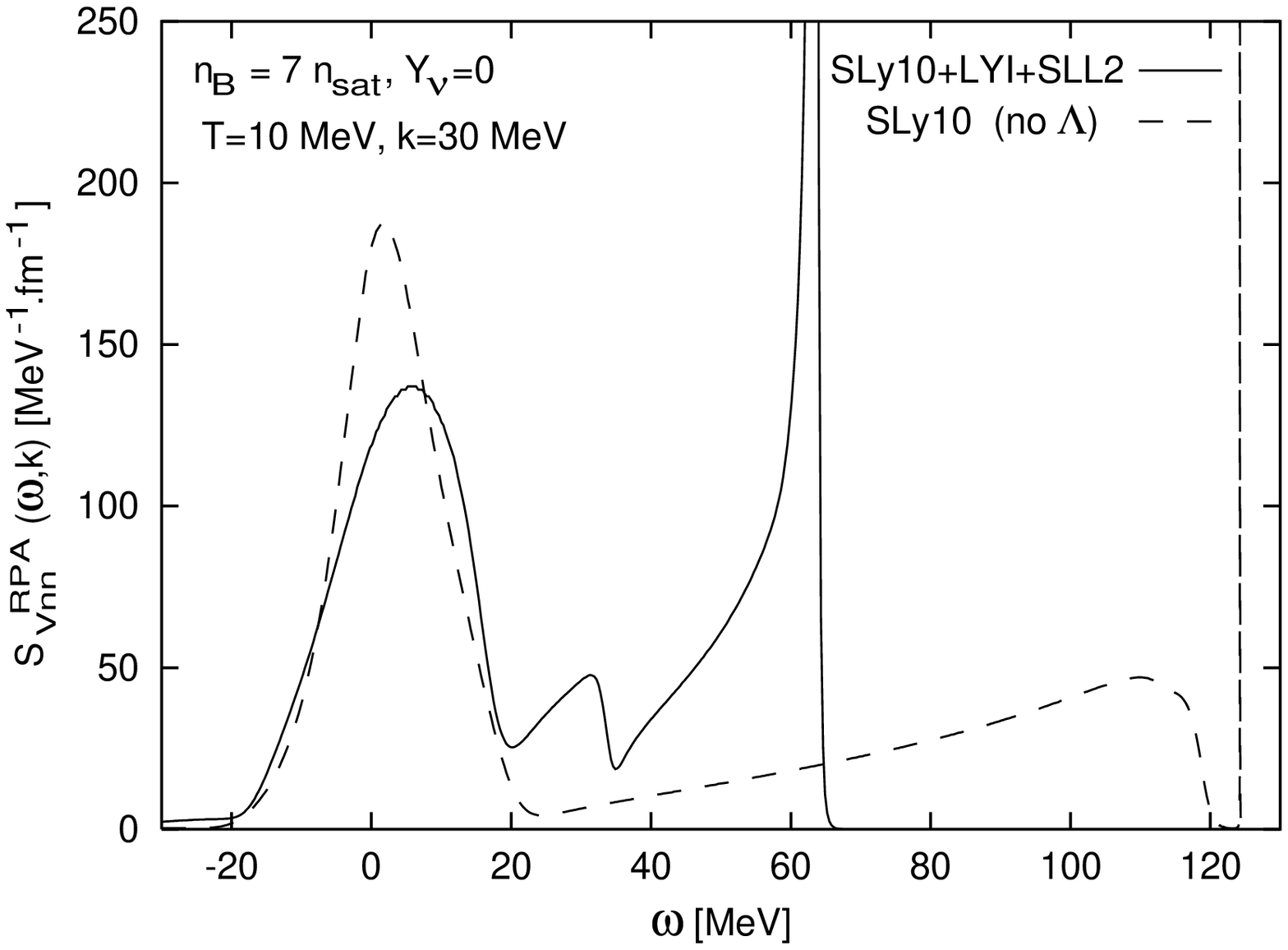,width=7.5cm}}
}
\vskip 0.2cm
\mbox{%
\parbox{16cm}{{\bf Fig. 7} -- Vector structure functions in
$npe\Lambda$ matter in beta equilibrium in the RPA approximation}
}
\end{figure} 

We observe that two such dips in $D_V$ develop in $npe$ matter in $\beta$ 
equilibrium whereas a third dip appears in $npe\Lambda$ matter. 
This translates in respectively two or three peaks in the vector response
function. The conditions of the right panel of Fig. 7 were chosen so that
these features are clearly visible. It is tempting to ascribe each dip 
to one species of baryons present in the system. As a matter of fact, 
the position of the dips is approximately given by the 
characteristic frequencies associated to each particle $\omega_{i}= k (k 
+ 2 p_{Fi})/(2 m_i^*)$. The deviation from the exact value is due to the 
fact that we have mixed states: the isospin is not a good quantum number, 
neither is the strangeness. For the density range relevant to neutron stars 
we only see the zero sound associated to the neutrons. 
The position ({\it i.e.} velocity) of the zero sound mode(s) depends
on the stiffness of the equation of state. On Fig. 7-b we see that, the
equation of state with hyperons being softer than in matter with only 
neutrons and protons, the zero sound mode propagates at smaller velocity. 
The trend is general, and the position of the zero sound peak for various 
choices of the NN, N$\Lambda$ and $\Lambda\Lambda$ forces follows their 
classification according to stiffness realized in \cite{M04a}:
SV+YBZ6 $>$ SkI3+YBZ6 $>$ SLy10+LYI.

\begin{figure}[htb]
\mbox{%
\parbox{7.5cm}{\epsfig{file=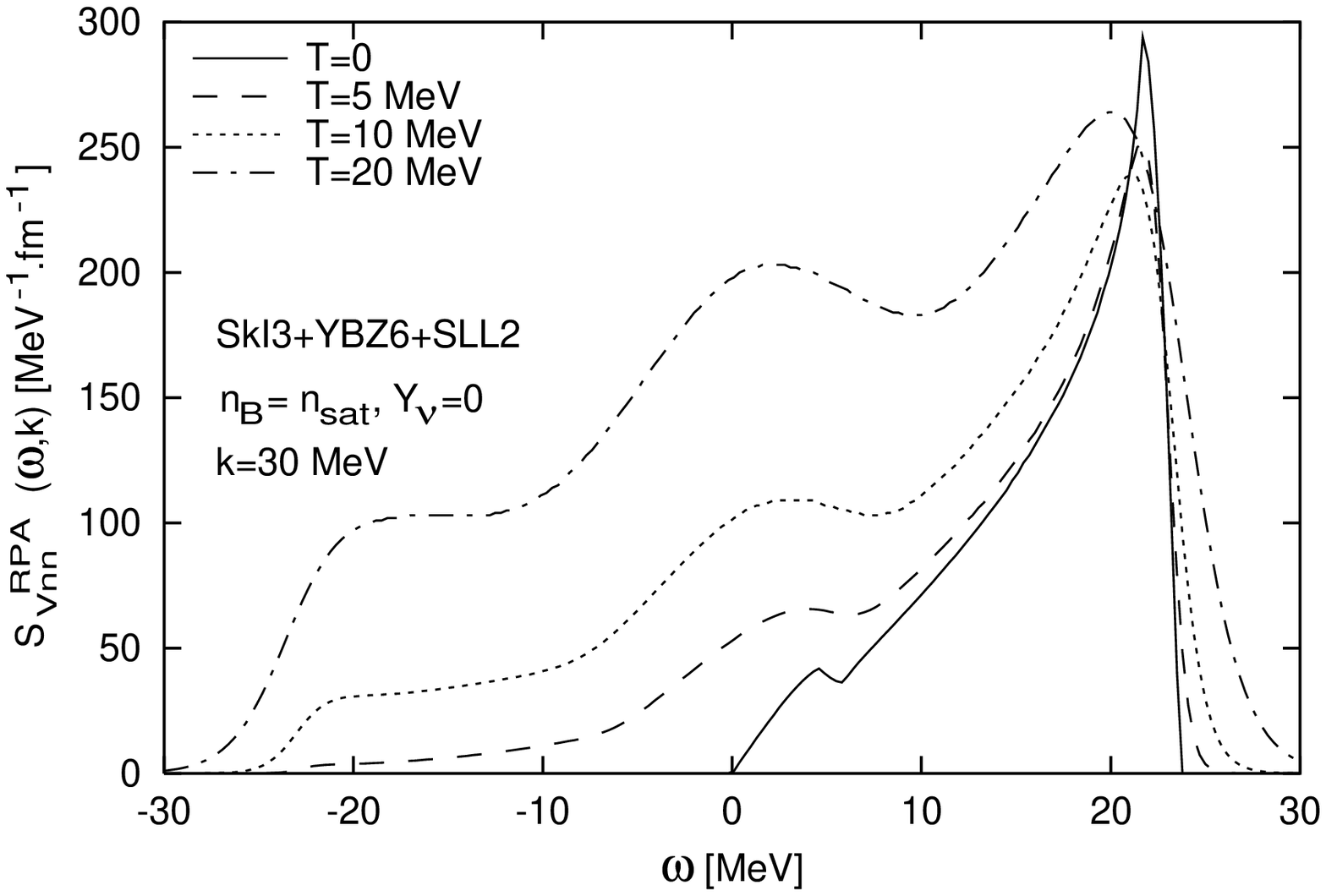,width=7.5cm}}
\parbox{1cm}{\phantom{aaaaa}}
\parbox{7.5cm}{\epsfig{file=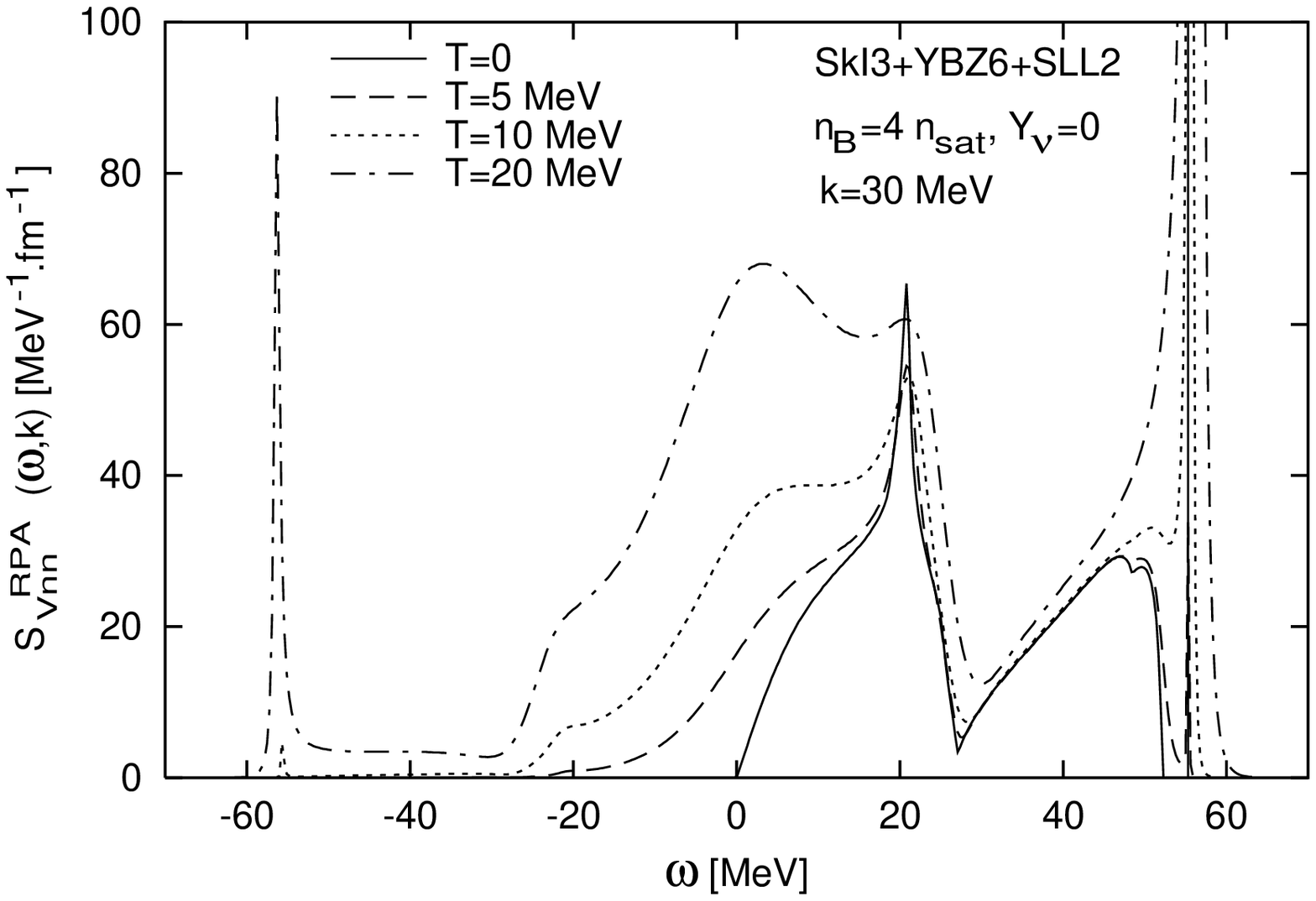,width=7.5cm}}
}
\vskip 0.2cm
\mbox{%
\parbox{16cm}{{\bf Fig. 8} -- Vector structure functions in
$npe\Lambda$ matter in beta equilibrium in the RPA approximation}
}
\end{figure} 

Fig. 7-a was drawn at zero temperature. When some temperature is applied to 
the system (see Fig. 8), the delta function (corresponding to an undamped
zero sound mode at T=0) merges with the remaining part of the structure 
function. As was already observed in the case of the axial response,
processes in which the neutrino gain energy in the collision ($\omega <0$) 
open as T increases.

\vskip 0.2cm
\par\noindent$\underline{\mbox{\sf c) Differential cross section}}$
\vskip 0.2cm

The differential cross section Eq. (\ref{eq:diffcross}) is displayed on 
Figure 9 at $n_B =2.5\ n_{\rm sat}$, {\it i.e.} shortly above the threshold
for hyperon formation. We compare the results for Skyrme models
SkI3+YBZ6+SLL2 and SV+YBZ6+SLL2.

Tiny secondary peaks are barely visible at high energy transfer
$\omega = \pm$ 60 MeV. This is what is left of the zero sound
contribution in the vector structure functions (see Figures 7 and 8)
after applying the Pauli blocking factor $1-f(E_\nu')$. Moreover,
in the chosen example $E_\nu=k=$ 30 MeV the condition $|\cos\theta|<1$ 
restrict the allowed energy ransfer $\omega$ to the  range 
$[-k - 2\, E_\nu-k]=[-30 - 30]$ MeV, so that the zero sound peak does 
not contribute alltogether.
 
\begin{figure}[htb]
\mbox{%
\parbox{7.5cm}{\epsfig{file=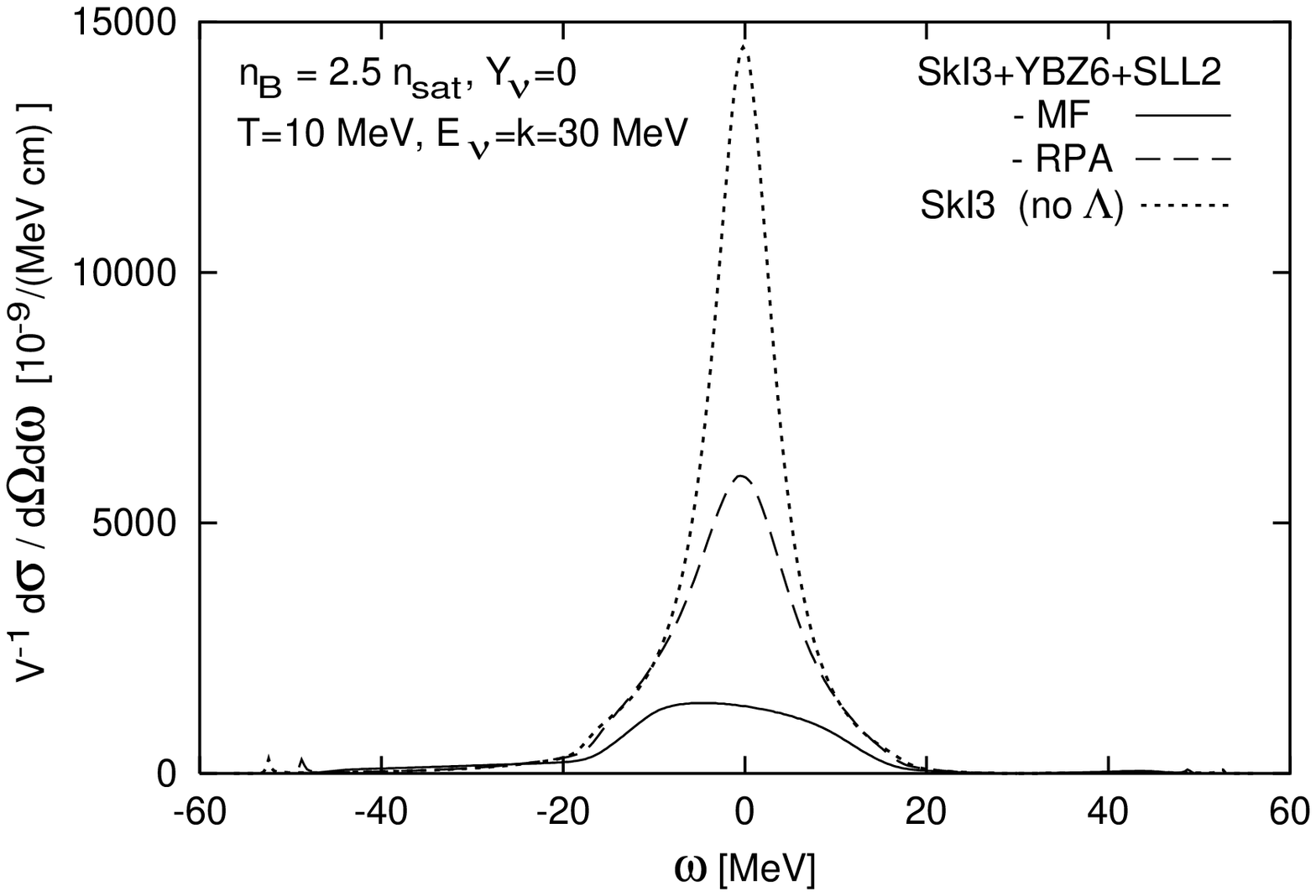,width=7.5cm}}
\parbox{1cm}{\phantom{aaaaaa}}
\parbox{7.5cm}{\epsfig{file=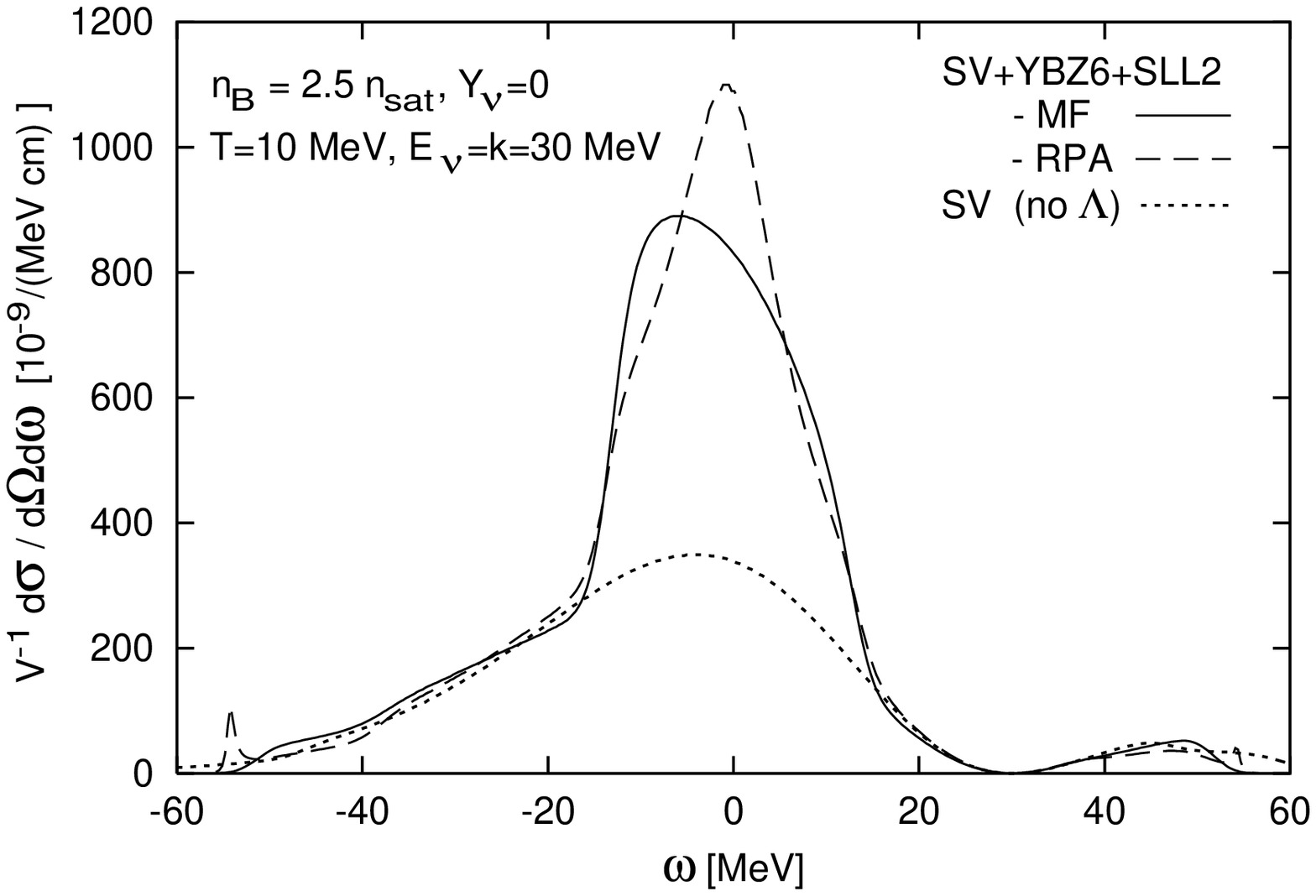,width=7.5cm}}
}
\vskip 0.2cm
\mbox{%
\parbox{16cm}{{\bf Fig. 9} -- Differential cross section in
$npe\Lambda$ matter in beta equilibrium in the RPA approximation}
}
\end{figure}

\subsection{Mean free path}
\label{S:mfp}

The mean free path is obtained by integration of the differential 
cross section. This parameter is relevant in the late stages of supernova 
explosion and first stages of proto neutron star cooling, when the 
neutrinos are still dynamically trapped in the hot and dense matter.
\beq
{1 \over \lambda} &=& \sigma = \int d\omega d\Omega \left( {d^2 \sigma 
\over d\omega d\Omega}\right) \nonumber \\
                  &=& {G_F^2 \over 4 \pi^2} \int_{-\infty}^{E_\nu} 
      d \omega {E_\nu' \over E_\nu} \int_{|\omega|}^{2 E_\nu-\omega} 
      q dq \left[1-f(E_\nu')\right]  \left[ (1-\cos\theta) S_V(\omega,q) 
      +(3+\cos\theta) S_A(\omega,q) \right] \nonumber \\
\eeq

\noindent
In the elastic limit where $S_V^{ij} \sim S_A^{ij} \sim S_0^{i} \delta^{ij}
\rightarrow 2 \pi n_i \delta(\omega)$ for $i,j \in \{ n,p,\Lambda \}$ 
and without Pauli blocking, this reduces to the textbook formula 
${1 \over \lambda} = \sum_{i=n,p,\Lambda} {1 \over \lambda_i} 
=\sum_{i=n,p,\Lambda} {G_F^2 \over \pi}  (c_{Vi}^2 + 3 c_{Ai}^2)\, 
E_\nu^2\, n_i $

\vskip 0.2cm
We studied the total cross section and the mean free path for both the
neutrino-free and trapped neutrino case. According to protoneutron star 
cooling calculations \cite{PRPLM99}, the characteristic energy of the 
neutrinos is $E_\nu \simeq 3 T$ in the former case whereas it is 
$E_\nu \simeq \mu_\nu$ in the latter case.

\begin{figure}[htb]
\mbox{%
\parbox{7.5cm}{\epsfig{file=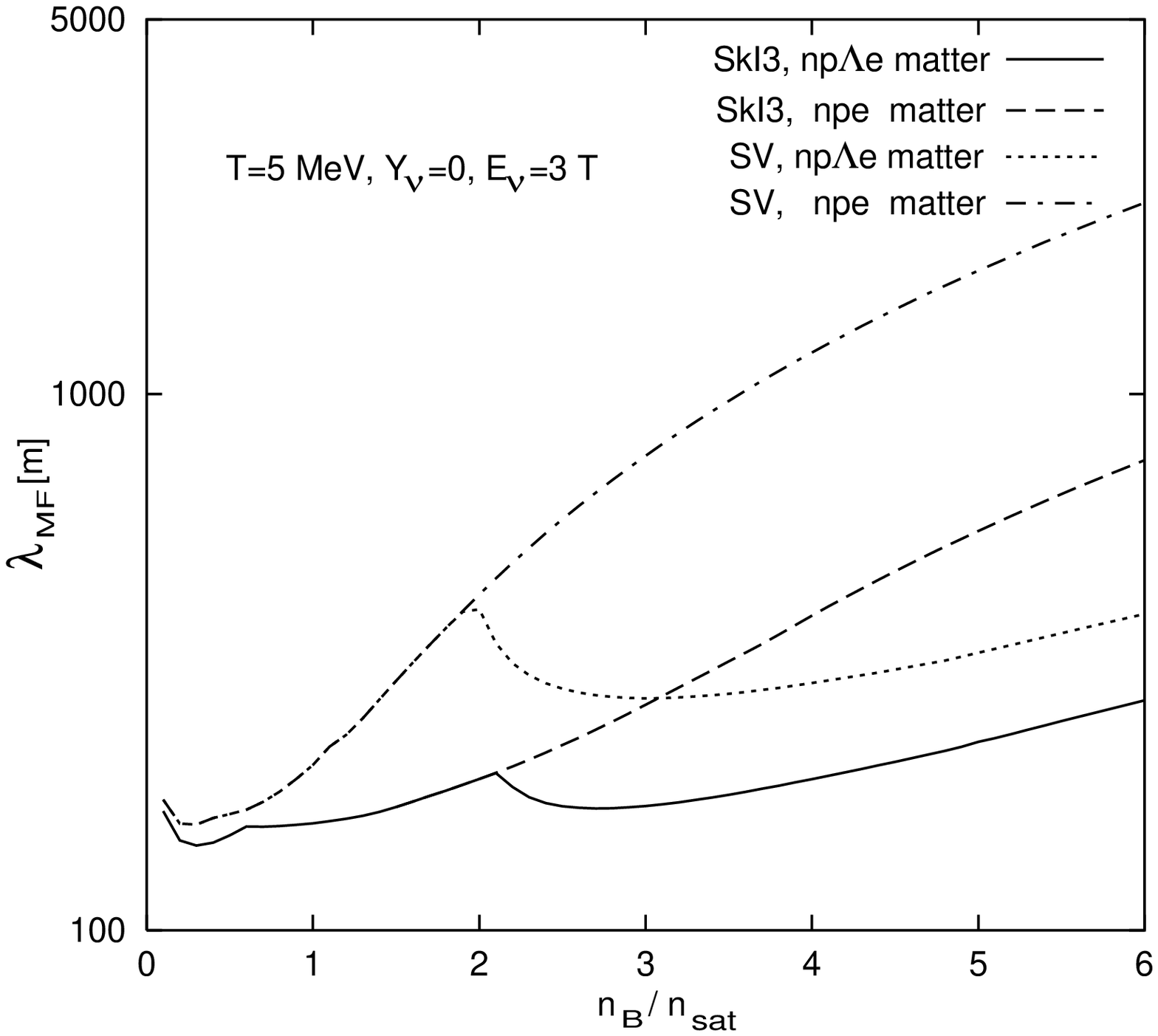,width=7.5cm}}
\parbox{1cm}{\phantom{aaaaa}}
\parbox{7.5cm}{\epsfig{file=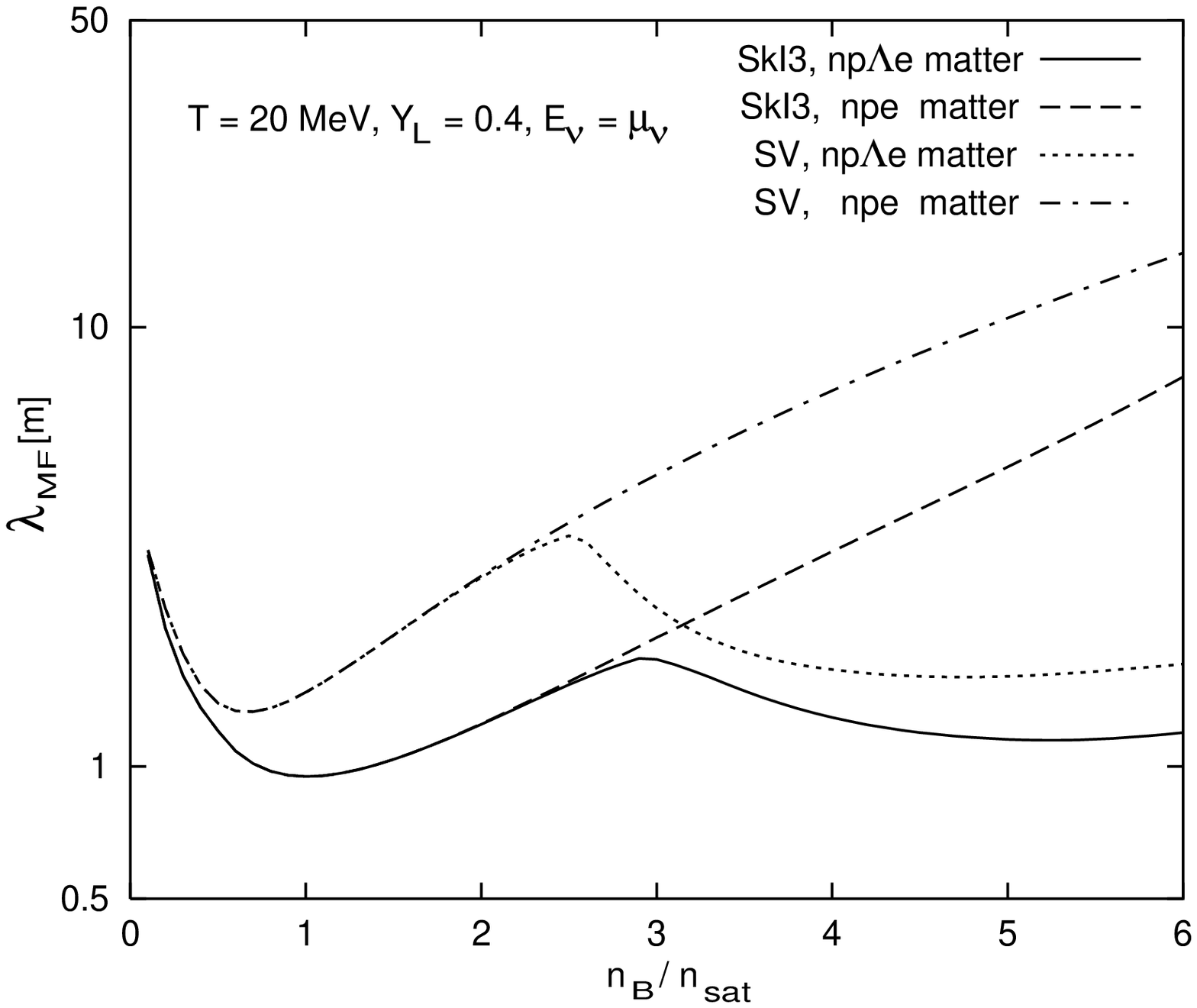,width=7.5cm}}
}
\vskip 0.2cm
\mbox{%
\parbox{16cm}{{\bf Fig. 10} -- Mean free path with and without hyperons
in the mean field approximation {\it (a)} In neutrino free neutron star
matter at $T$=5 MeV {\it (b)} In protoneutron star matter with trapped 
neutrinos at T=20 MeV, $Y_L$=0.4}
}
\end{figure} 

Fig. 10 shows the mean free path in the mean field approximation.
On the left panel we have represented the neutrino-free case $Y_\nu=0$
for a moderate temperature $T=5$ MeV while the right panel displays
the case of trapped neutrinos $Y_L=0.4$ with a higher temperature $T=20$ 
MeV. It can be seen that, at the mean field approximation, the total cross 
section is enhanced by the apparition of hyperons and the mean free path 
correspondingly reduced. The same behavior was observed by Reddy {\it et al.}
\cite{RPL98} and  explained by the combined effect of the modification of 
the chemical potentials of the nucleons and the opening of new 
scattering channels when new degree of freedoms are available. The results 
of the SkI3+YBZ6+SLL2 and SV+YBZ6+SLL2 are similar, with the mean free path
being larger for the SV parametrization corresponding to lower effective 
masses: At the mean field level, $\sigma \propto {\cal I}{\rm m} \Pi_0 
\propto m_*^2$ (see Eq. (\ref{eq:ImPi})) and $\lambda \sim 1/\sigma$.

The mean free path is observed to increase with higher densities. This
behavior was also found in nonrelativistic models by Reddy {\it et al}
\cite{RPLP99}, whereas relativistic models yield a decreasing mean free 
path with increasing density. The discrepancy was explained by  Reddy 
{\it et al} \cite{RPLP99} as the result of two competing mechanisms, 
{\it i.e.} the decreasing of the effective masses (see previous alinea)
and the increase of the chemical potentials: in the nonrelativistic
Skyrme model the decreasing $m_*$ is the dominant effect.

\begin{figure}[htb]
\mbox{%
\parbox{7.5cm}{\epsfig{file=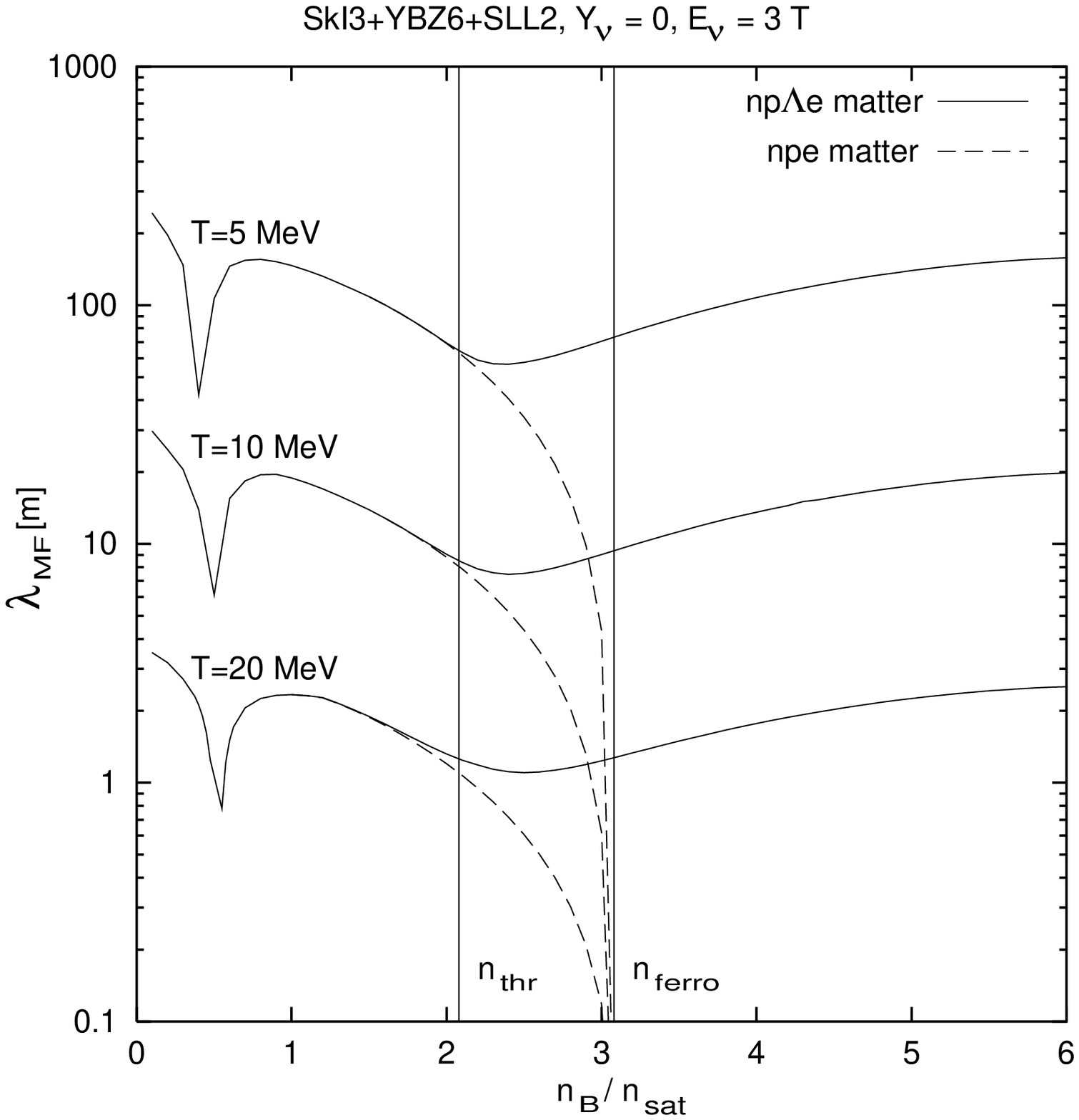,width=7.5cm}}
\parbox{1cm}{\phantom{aaaaa}}
\parbox{7.5cm}{\epsfig{file=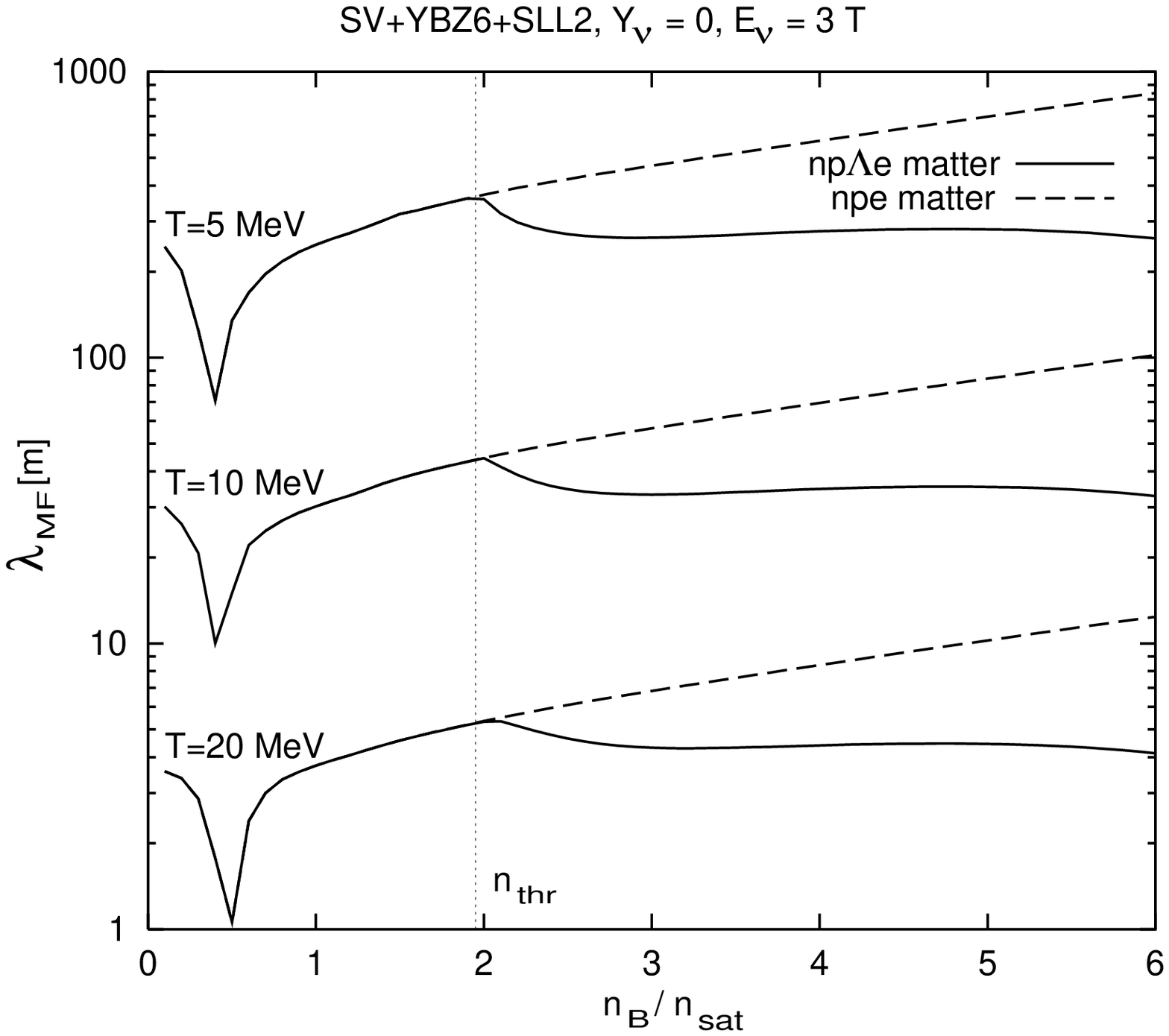,width=7.5cm}}
}
\mbox{%
\parbox{7.5cm}{\epsfig{file=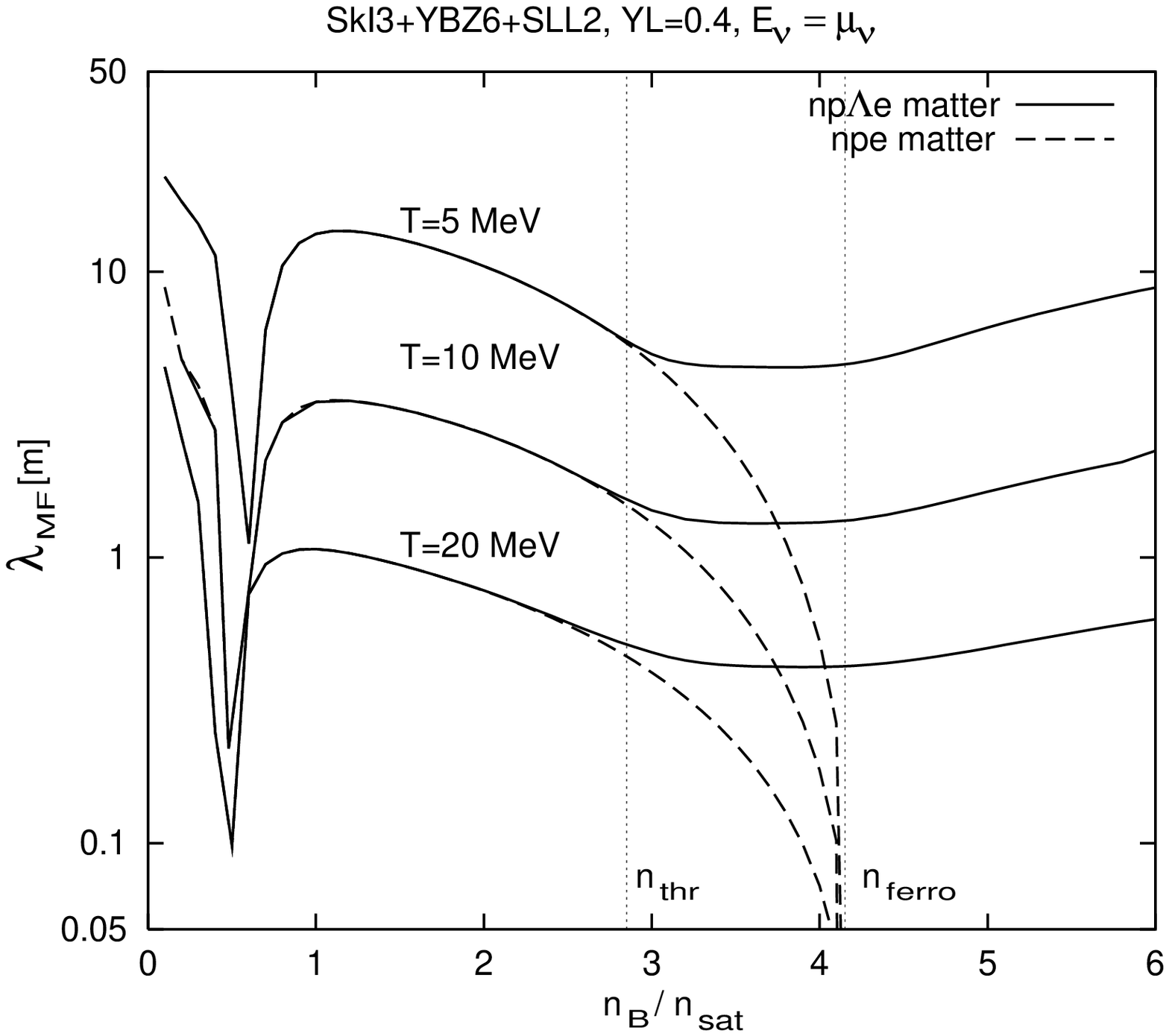,width=7.5cm}}
\parbox{1cm}{\phantom{aaaaa}}
\parbox{7.5cm}{\epsfig{file=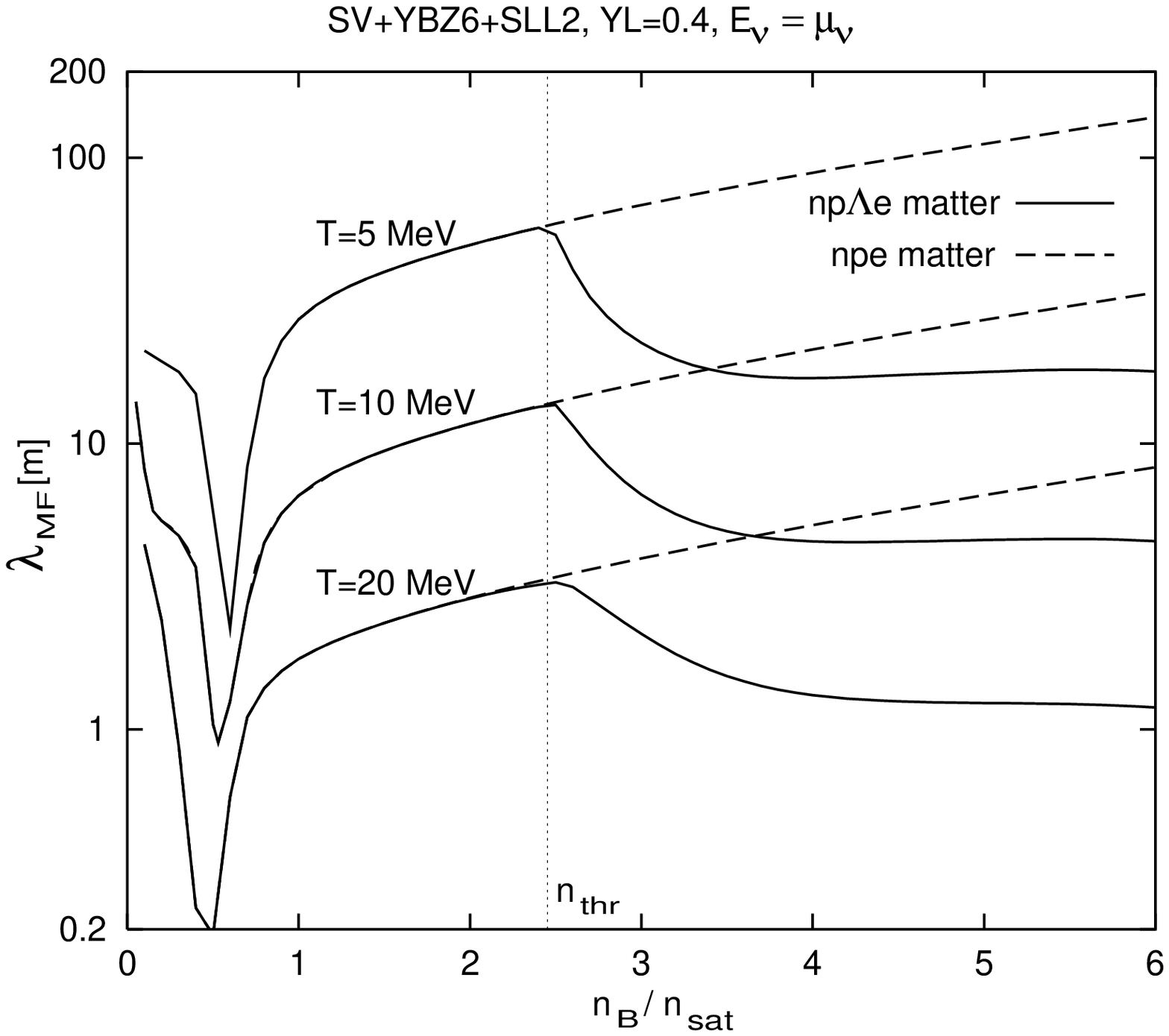,width=7.5cm}}
}
\vskip 0.2cm
\mbox{%
\parbox{16cm}{{\bf Fig. 11} -- Mean free path $\lambda_{RPA}$ with 
and without hyperons in the random phase approximation. On the left 
hand side  $\lambda_{RPA}$ was calculated with the SkI3+YBZ6+SLL2 
parametrization, on the right hand side with the SV+YBZ6+SLL2
parametrization. Top panels: neutrino free neutron star matter. 
Bottom panels: protoneutron star matter with trapped neutrinos ($Y_L=0.4$)}
}
\end{figure} 

The mean free path in the random phase approximation is displayed in
Fig. 11 in neutrino-free matter on the top panels and with trapped
neutrinos on the bottom panels for parametrizations SkI3+YBZ6+SLL2
(left) and SV+YBZ6+SLL2 (right).

We first note an enhancement of the cross section at subnuclear density
for both parametrizations. This is related to the peak present in the vector 
response function which was noticed in the previous section, and signals 
the mechanical instability to non homogeneous "pasta" phases. Our model 
is not valid any more in this density range; a calculation would need to 
be performed along the lines of reference {\it e.g.} \cite{HPGP04}. 

The dramatic enhancement of the cross section in the RPA approximation due 
to the onset of ferromagnetism which was reported in several previous 
works \cite{RPLP99,NHV99,M01} is also observed with parameter set SkI3 
in $npe$ matter. It is indicated on the figure by the line $n_{\rm ferro}$. 
The mean free path in $npe$ matter goes to zero at this point. When 
hyperons are present in the system, the ferromagnetic transition can be 
avoided when the threshold density for hyperon formation is lower than 
the critical density for ferromagnetism in npe matter \cite{M04a}. For the 
set SkI3+YBZ6+SLL2 we have $n_{\rm thresh}^\Lambda= 2.08\ n_{\rm sat}$ 
while in npe matter $n_{\rm ferro}=3.08\ n_{\rm sat}$. As a consequence, 
the pole is avoided and the mean free path goes on gently increasing.

The parameter set SV which was selected as the only Skyrme model in which 
the ferromagnetic transition does not appear, permits us to study the
influence of the RPA corrections in the absence of the pole. We see 
on the right panels of Fig. 11 that the mean free path behaves in this case
in a way similar to the mean field result at densities $n_B > n_{\rm sat}$.

Fig. 12 shows the ratio of the total cross sections 
$\sigma_{\rm RPA}/\sigma_{\rm MF}= \lambda_{\rm MF}/lambda_{\rm RPA}$ 
in the RPA and mean field approximation.
A residual enhancement in the vicinity of the ferromagnetic instability 
remaining from the np sector prior to the hyperon formation threshold is 
visible in the SkI3+YBZ6+SLL2 parametrization. In the SV+YBZ6+SLL2 
parametrization the cross section is somewhat enhanced with respect to
the mean field result: even in the absence of ferromagnetism, this is 
still at variance with the relativistic prediction \cite{RPLP99,YT99,MP02}.

\begin{figure}[htb]
\mbox{%
\parbox{7.5cm}{\epsfig{file=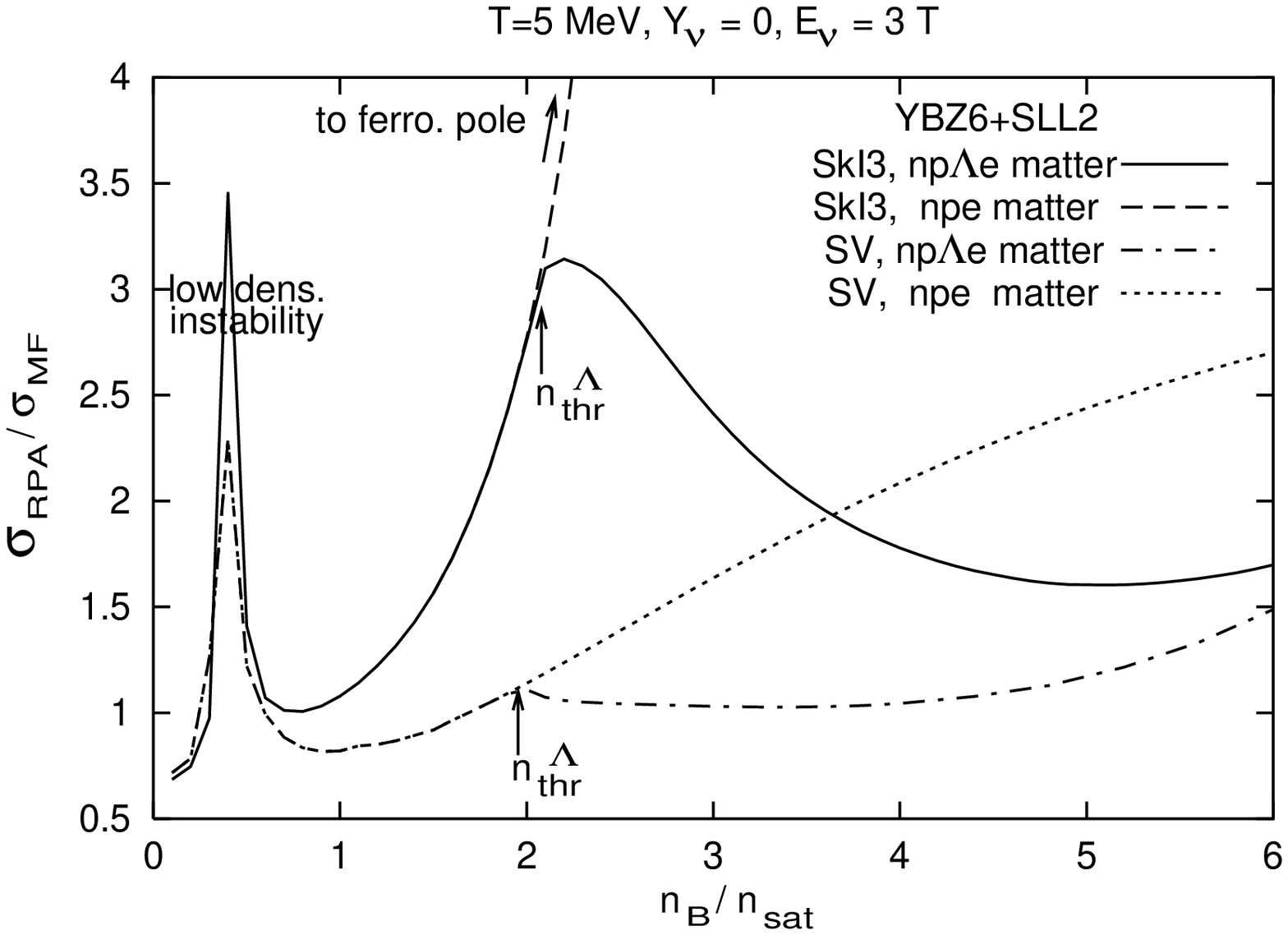,width=7.5cm}}
\parbox{1cm}{\phantom{aaaaa}}
\parbox{7.5cm}{\epsfig{file=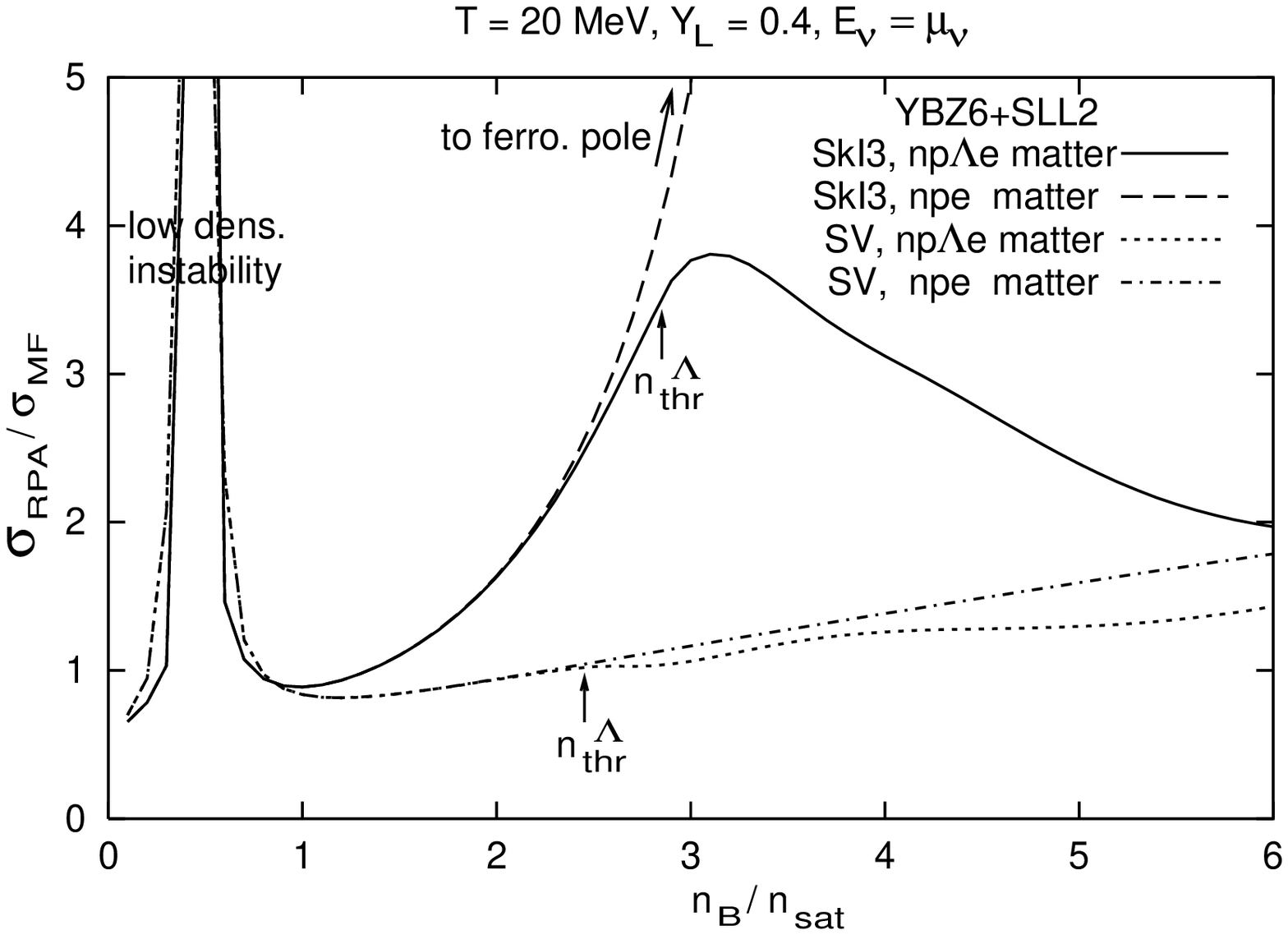,width=7.5cm}}
}
\vskip 0.2cm
\mbox{%
\parbox{16cm}{{\bf Fig. 12} -- Enhancement factor for cross sections
calculated in the random phase approximation with respect to mean field 
approximation {\it (a)} In neutrino free neutron star matter at $T$=5 MeV 
{\it (b)} In protoneutron star matter with trapped neutrinos at T=20 MeV,
$Y_L$=0.4}
}
\end{figure} 

We finally explored the impact of the neglect of the other hyperons
$\Sigma, \Xi$ by performing a mean field calculation in the model
of Balberg and Gal \cite{BG97} and another nonrelativistic model
based on the Seyler-Blanchard effective interaction by Banik and 
Bandhyopadhyay \cite{BB00}. Both models include all species of hyperons.
The calculation was nevertheless performed here in a reduced version of
these models with only the $\Sigma^-$ and $\Lambda$ hyperons taken
into account and no muons. The $\Sigma^-$ hyperons appear at 
$1.87\ n_{\rm sat}$ and the $\Lambda$  at $2.32\ n_{\rm sat}$  
in the model of Balberg and Gal with the parameter set corresponding 
to the choice $\gamma=4/3$, while the  $\Sigma^-$ are formed at
$1.52\ n_{\rm sat}$ and the $\Lambda$ at $2.82\ n_{\rm sat}$ in
the model of Banik and Bandhyopadhyay. The mean free path obtained with
such models is shown in fig. 13 in neutrino free matter at $T=10$ MeV 
for a neutrino of energy $E_\nu=30\ MeV$. The reduction of the mean free
path at the formation threshold of a new hyperon is also observed
in these models. The effect is less visible for the second hyperon than 
for the first. The behavior of the mean free path obtained from the model
of Banik and Bandhyopadhyay is qualitatively similar to that resulting
from the Skyrme parametrizations considered in the present work.
The decrease of the mean free path with increasing density obtained 
from the model of Balberg and Gal is easily understood from the fact 
that whereas both the Skyrme and the Seyler-Blanchard parametrizations 
have decreasing effective masses, Balberg and Gal keep theirs fixed 
at the free value (see the discussion below Fig. 10).

\begin{figure}[htb]
\mbox{%
\parbox{7.5cm}{\epsfig{file=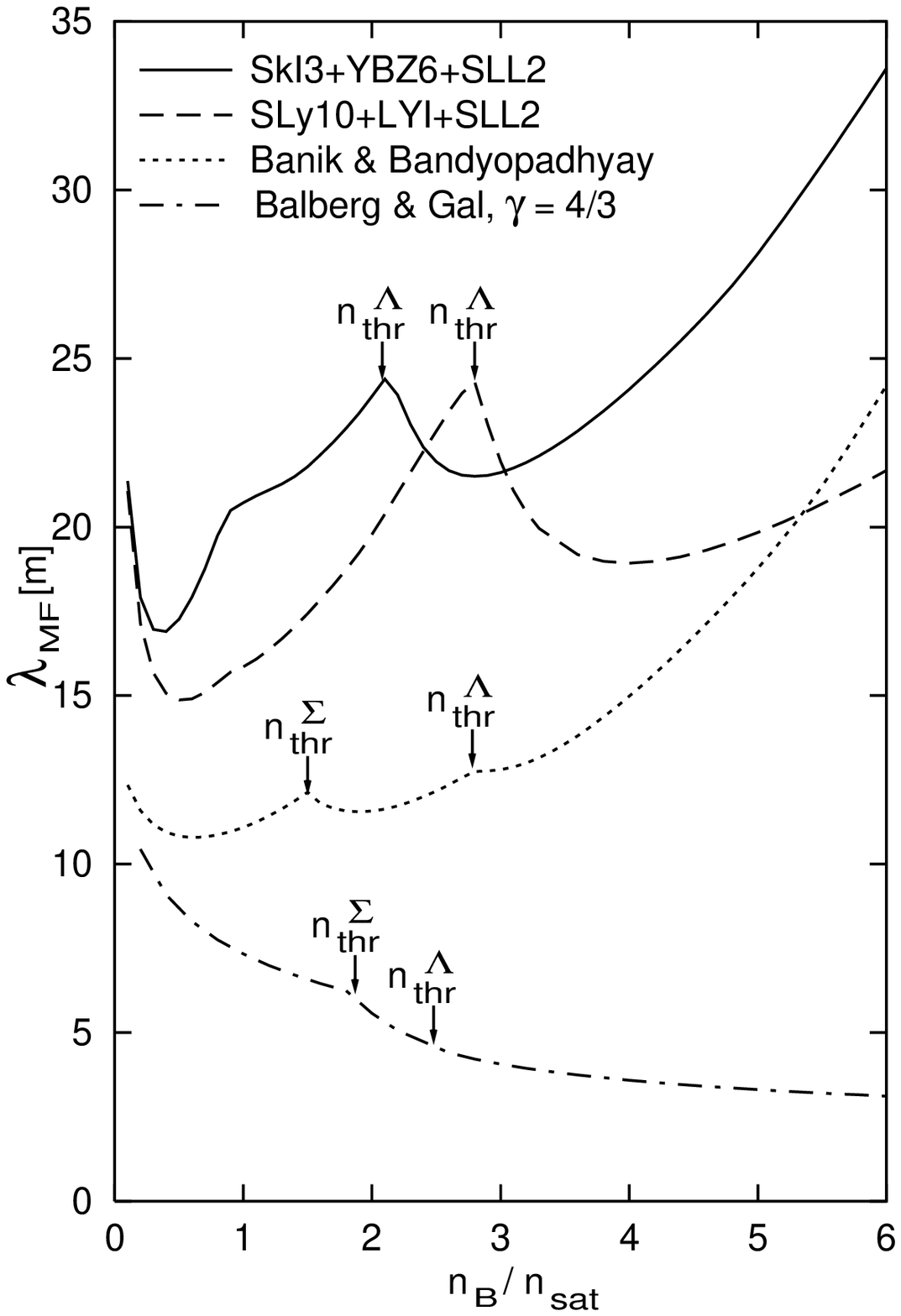,width=7.5cm}}
\parbox{0.5cm}{\phantom{aaa}}
\parbox{7.5cm}{{\bf Fig. 13} -- Mean free path in the mean field approximation
at $T=10$ MeV, $E_\nu=3\ T$, $Y_\nu=0$  for various nonrelativistic models 
of baryonic matter}
}
\end{figure}

\section{Conclusion}
\label{S:concl}

We have performed for the first time a calculation of the neutrino-baryon 
scattering rates including the hyperons at the RPA level in a nonrelativistic
model. The nuclear interactions were described by Skyrme forces calibrated
to reproduce the available data on nuclei and hypernuclei and tested for
their suitability to describe the properties of neutron stars. 

At the mean field level the hyperons open new scattering channels and tend 
to decrease the mean free path as compared to its value in $npe$ matter. In
the random phase approximation the cross section is very sensitive to the
degree of stability of the matter with respect to density or spin density 
fluctuations. The hyperons help delaying or removing the onset of a 
ferromagnetic instability usually present in Skyrme models. The drastic
reduction of the mean free path at the transition
is thus avoided. A residual enhancement of the cross section nevertheless
remains in most cases.

Even in the complete absence of a ferromagnetic instability as provided by 
the parametrization SV+YBZ6+SLL2, we still obtained results which are at 
variance with the prediction of relativistic models: {\it (i)} The
mean free path increases with density. This may be ascribed to a rapid
decrease of the effective masses of the baryons in these models. This
feature was observed previously in $npe$ matter \cite{RPLP99}. {\it (ii)}
The RPA cross section was still slightly enhanced with respect to the mean
field value. This indicates a less repulsive spin interaction as density 
increases.

\vskip 0.2cm
Our model is still very schematic. Several items would need to be improved 
in future work, among which:
 
{\it (i)}  {\sf Particle-hole interaction} 
\par\noindent
The Skyrme interaction was used for convenience in order
to keep the model as simple as possible. While it was carefully
selected in order to avoid known problems at high density such as the 
behavior of asymmetry energy, causality and onset of ferromagnetism, 
it is still being driven at the limit of its reliability. The behavior 
of the interaction in the spin-spin channel still remains unclear.
An other issue is the description of the $\Lambda$-$\Lambda$ force 
which plays an important role at high density where the hyperon fraction
is large, whereas the description of the hyperon-hyperon interaction in
phenomenological formalisms is still very poor. 

Both points could be substantially improved while still keeping the same 
level of simplicity for the Dyson equations and neutrino-baryon scattering
cross sections, if we could obtain a parametrization of the polarized 
energy density functional and effective masses from microscopical variational 
or Brueckner-Hartree-Fock calculations. Landau parameters extracted from 
BHF calculations have indeed been used very recently to calculate the neutrino
scattering mean free path in neutron matter with the Nijmegen NSC97e potential
\cite{MVB03} and for $npe$ matter in $\beta$ equilibrium with the Argonne 
AV${}_{18}$ potential plus 3-body forces \cite{SLGZ03}. The spin-spin 
interaction was actually found to be increasingly attractive by both groups,
leading to a strongly reduced cross section in the random phase approximation.

Brueckner-Hartree-Fock calculations also offer a coherent framework to 
calculate the hyperon-hyperon interaction in medium starting from bare 
baryon-baryon potentials and then cross checking the results with the 
data on double hypernuclei.  The parametrization of Brueckner results 
for matter with an hyperonic component is currently under way and will 
be the subject of another paper \cite{ournext}.

{\it (ii)} {\sf Full RPA} 
\par\noindent We limited ourselves to the monopolar $l=0$ 
Landau approximation. As previous studies in pure neutron matter or
symmetric nuclear matter have shown \cite{M01,HNP97,BVA95} the full 
RPA approximation can represent an appreciable correction to the simple 
$l=0$ result. While the extension of the full RPA to np$\Lambda$e
matter should not present any conceptual difficulties, the additional
amount of work required is not yet worthwhile, considering the other 
uncertainties of the model.

{\it (iii)} {\sf Relativistic effects} 
\par\indent 
The main advantage of the nonrelativistic model is that the Dyson equation 
has a simpler structure, especially if we restrict the calculation to the 
monopolar approximation. This allows to explore a greater variety of
many-body effects and chemical composition of the matter. At the high 
densities considered in protoneutron stars, a relativistic model however 
appears to be a desirable requirement.

{\it (iv)} {\sf Charged current processes} 
\par\indent
The present work explored the role of hyperons in neutrino processes
on the example of the scattering through the neutral current. The same 
methods as used here can be applied to the charged current process.
In protoneutron stars, the charged current process $ \nu_e +n 
\rightleftharpoons p + e^-$ is actually dominant. In the subsequent
long cooling phase, the triangle rule for conservation of momentum
constrained to the immediate vicinity of the Fermi surface is less
restrictive in the direct URCA process involving hyperons such
as $p + e^- \rightleftharpoons \Lambda + \nu_e$, which makes this an 
interesting mechanism for an efficient cooling of the star \cite{NScool-hyp}.

\vskip 0.5cm
\noindent {\large{\bf Acknowledgements}}
\vskip 0.2cm

This work was supported  by the spanish-european (FICYT/FEDER) 
grant number  PB02-076. Part of it was realized during a stay at the 
Departament d'Estructura i Constituents de la Materia of Barcelona 
University. Several discussions with A. Polls are gratefully acknowlegded.

\end{document}